# Novel Bribery Mining Attacks in the Bitcoin System and the "Bribery Miner's Dilemma"

Junjie Hu
Department of Computer Science
University of Electronic Science and
Technology of China
Chengdu, Sichuan, China
hujj@std.uestc.edu.cn

Chunxiang Xu
Department of Cyberspace Security
University of Electronic Science and
Technology of China
Chengdu, Sichuan, China
chxxu@uestc.edu.cn

Zhe Jiang
Department of Mathematical Sciences
University of Electronic Science and
Technology of China
Chengdu, Sichuan, China
zhejiang@std.uestc.edu.cn

Jiwu Cao
Department of Electronic Engineering
University of Electronic Science and
Technology of China
Chengdu, Sichuan, China
2021190504024@std.uestc.edu.cn

**ABSTRACT**

Mining attacks allow adversaries to obtain a disproportionate share of the mining reward by deviating from the honest mining strategy in the Bitcoin system. Among them, the most well-known are selfish mining ($SM$), block withholding ($BWH$), fork after withholding ($FAW$) and bribery mining. In this paper, we propose two novel mining attacks: bribery semi-selfish mining ($BSSM$) and bribery stubborn mining ($BSM$). Both of them can increase the relative extra reward of the adversary and will make the target bribery miners suffer from the "bribery miner dilemma". All targets earn less under the Nash equilibrium. For each target, their local optimal strategy is to accept the bribes. However, they will suffer losses, comparing with denying the bribes. Furthermore, for all targets, their global optimal strategy is to deny the bribes. Quantitative analysis and simulation have been verified our theoretical analysis. We propose practical measures to mitigate more advanced mining attack strategies based on bribery mining, and provide new ideas for addressing bribery mining attacks in the future. However, how to completely and effectively prevent these attacks is still needed on further research.

## 1 INTRODUCTION

Bitcoin [1] is a decentralized cryptocurrency based on blockchain technology, which is proposed by Satoshi Nakamoto in November 2008. Unlike most currencies, Bitcoin does not rely on specific currency institutions for issuance. It is generated through a large amount of calculations according to specific algorithms. The Bitcoin system uses a distributed database composed of numerous nodes in the entire peer-to-peer network to confirm and record all transactions, and adopts cryptographic design to ensure the security of the whole process of currency circulation. The decentralization of peer-to-peer network and consensus algorithms can ensure that currency value cannot be artificially manipulated through the mass creation of Bitcoin. Cryptographic-based designs allow Bitcoin only to be transferred or paid for by real owners.

In the Bitcoin system, participants (miners) can get rewards by adding transaction records to the ledger (blockchain), which requires miners to solve cryptographic puzzles as a proof of work (PoW) [36]. The first miner to solve the puzzle and generate a valid block can obtain block rewards (6.25 Bitcoins in 2023). The process of miners solving cryptographic puzzles and generating blocks is called "mining process". When two or more blocks are generated and published simultaneously in the system (due to network communication delay), forking occurs. To maintain consistency, one of the brunches will be selected by the system and eventually become the main chain. Once miners on other branches receive the longest chain, they will shift their attention and mining power to the main chain. In the Bitcoin system, the difficulty of solving cryptographic puzzle is adjusted per two weeks to maintain the average generation time of blocks as a constant (10 minutes). However, due to the current mining power's hash rate exceeding $3.3 \times 10^{20}$ Hash/s [35], it probably takes a single miner several months or even years to solve a password puzzle [2]. Therefore, to attain stable income, miners tend to unit to form a miner pool. Most mining pools have a pool manager responsible for assigning work and rewards. When a mining pool finds a block, the miners in the mining pool will share rewards in terms of their contributions (the number of shares submitted).

Since cryptocurrencies have monetary value, they naturally become a valuable target for attack. Although the design of Bitcoin ensures security, previous studies have shown that adversaries can increase their rewards when deviating from honest mining strategies, such as selfish mining [8], block with holding ($BWH$) [20], fork after withholding ($FAW$) [24], and bribery attacks [26]. In selfish mining attacks, adversaries intentionally hide discovered blocks to form a private chain and continue to mine on the private chain. When a block is generated on the public chain, adversaries selectively publish blocks on the private chain, and get disproportionate rewards by wasting the mining power of honest miners. Semi-selfish mining ($SSM$) [18] is a mining strategy constructed on the basis of $SM$ which divides mining power into two parts. The consumptions of two parts of mining power are similar to selfish and honest pools, respectively. Most of mining power is applied to mining on the private chain while the other small portion is utilized to mine on public chain. The design of $SSM$ can significantly reduce the system forking rate while only slightly reducing the profit of selfish miners. Briefly, $SSM$ can balance benefit and forking rate. In the $BWH$ attack, the adversaries divides their mining power into innocent pool and infiltration pool. When infiltration pool finds a valid block (full proof of work, FPoW), he withholds it and continues to submit other shares (partial proof of work, PPoW) to obtain the share reward. [20] has shown that $BWH$ attacks are more profitable than honest mining ($HM$) when adversaries segment their mining power appropriately. However, when two pools use $BWH$ attacks against each other (both pools have lower reward than $HM$), they will encounter the "miner's dilemma". The design principle of $FAW$ attack is similar to $BWH$ attack. More specifically, the only difference is that in $BWH$ attack, the adversary will discard the discovered FPoW, while in $FAW$, the attacker will reserve this FPoW. When other miners (not in the victim pool) find a valid block, the adversary will release and submit the previously reserved FPoW, causing a fork (similar to $SM$) to win in the forking competition and obtain share reward. Compared with $BWH$, $FAW$ can get more reward while avoiding the miner's dilemma. In bribery mining attack, once forking occurs, the adversary will try to win in the forking competition by bribing part of honest miners (called target bribery pool) to extend its branch and paying the bribe to the target bribery pool to obtain higher profits.

In this paper, we propose two novel strategies of mining attack to increase the reward of the adversary. Moreover, We model multi-target bribery pools and prove target pools would suffer "the bribery miner's dilemma" in $BSSM$ and $BSM$. Finally, we put forward practical measures to mitigate the high-level attacks based on bribery mining. However, how to prevent such attacks completely remains an unresolved issue.

We summarize our contributions as follows:
- Adversaries can get higher reward through bribery attacks in semi-selfish mining attack and stubborn mining attack. We discussed the situation where adversaries launch bribery attacks. In a forking competition situation, adversaries can bribe other honest miners to extend the attacker's branch, increasing the probability of successful forking competition and hence obtaining higher profits.

- We further proposed bribery semi-selfish mining ($BSSM$) and bribery stubborn mining ($BSM$). $BSSM$ combines bribery mining and $SSM$. Simulation experiment results indicate that $BSSM$ can result in 6% relative extra reward for adversaries in comparison with $SSM$ with the same chain growth rate.
- The target bribery pools will suffer the "briery miner's dilemma" in $BSSM$ and $BSM$ under the multi-target bribery pool model. On the one hand, from the perspective of each target bribery pool, his optimal strategy is to accept bribes and extend attacker's branch. However, he will suffer losses if all target bribery pools reject bribes. On the other hand, from the standpoint of target bribery pools, their optimal strategy is to reject bribes.
- We proposed practical countermeasures to mitigate higher-level bribery attacks, and provided new ideas for mitigating bribery mining in the future.

## 2 PRELIMINARIES
### 2.1 Bitcoin Background

**Mining Process.** The issuance process of Bitcoin is implemented by the Bitcoin system generating a certain number of Bitcoins as rewards for miners, in which miners play the role of currency issuers. The process of generating new blocks is also known as mining. All Bitcoin transactions need to be packaged into blocks and recorded in the ledger. The miner who first finds the nonce that meets the difficulty requirements can get the coinbase reward. The mining process motivates miners to maintain the security of blockchain. The total number of bitcoins was initially set to 21 million. Each miner who publishes a block can get 50 Bitcoins as a coinbase reward initially, which halves per 4 years. It is expected that the coinbase reward will no longer be able to be further subdivided until 2104, which results in completing the issuance of all Bitcoins.

**Forks.** When multiple miners broadcast the blocks discovered by them simultaneously, blockchain forking occurs, since other miners will consider the first received valid block as the header [33]. One branch will compete successfully thus becoming the main chain eventually. Miners who publish blocks on the main chain will obtain corresponding coinbase rewards, while others will not get any rewards. Note that forks may also occur intentionally, such as $SM$ attack [8] or $FAW$ attack [24].

**Mining Pool.** With the increasing investment of mining power in Bitcoin, the probability of a miner discovering a valis block becomes extremely small. Nowadays, miners tend to participate in an organization called mining pool. In general, a mining pool consists of a pool manager and multiple peer miners. All participants collaborate to solve the same cryptographic puzzle. Once the mining pool generates a valid block successfully, participants will share rewards according to the distribution protocol, such as Pay Per Share (PPS), Pay Per Last N Shares (PPLNS), Pay Proportionally (PROP) [3] and so on. In theory, the rewards of miners are proportional to their mining power directly. Therefore, miners who participate to the mining pool can reduce the difference in profits significantly. Currently, most of the blocks in Bitcoin are generated by mining pools, such as AntPool [4], Poolin [5], and F2Pool [6].

### 2.2 Related Work

**Selfish Mining.** Attackers can generate a fork through selfish mining ($SM$) intentionally to obtain additional rewards [7,8]. Specifically, in $SM$ attack, adversaries hide discovered blocks intentionally, forming a private chain and continuing to extend it. Once a new valid block is generated in public chain, attackers selectively publish blocks on the private chain, and obtain disproportionate rewards by wasting the mining power of honest miners. It is expected that the motivation to mine will rely more on transaction fees rather than block rewards due to the continuous decline in coinbase rewards. Once the transaction volume of Bitcoin decreases, these transaction fees will not be enough to compensate miners for their investment in computing resources. Consequently, some miners may stop mining temporarily, which will threaten the security of Bitcoin system. [9] introduces the incentive mechanism of Bitcoin when the total computing power of the system decrease. [16] expands the underlying model of $SM$ attack, further optimizes the upper bound of optimal strategy rewards, and lowers the minimum threshold for obtaining extra returns from $SM$. [17] supplements the action space of $SM$, models as Markov Decision Process (MDP), and pioneers a new technology to solve the nonlinear objective function of MDP, resulting in a more powerful $SM$ strategy. Under the same assumption, relevant studies conduct a series of discussions on the mining strategies of rational mining pools [10,11,12,13]. [14] provides some simulation results when involving multiple independent selfish mining pools or stubborn mining pools. [15] theoretically studies the equilibrium of multiple independent selfish mining pools. [37] focuses on the classic selfish mining attacks in the blockchain, explores the strategies to deal with the attacks from the perspective of game theory, and further depicts the equilibria state of the system under the competition of various strategies. However, due to the high forking rate caused by $SM$, these attacks are not practical. Once honest miners discover abnormal forking rate, they may exit the blockchain system. $SM$ attack is no longer meaningful with the departure of honest miners. [18] proposes semi-selfish mining ($SSM$) attack, which can achieve a balance between revenue and forking rate. [19] proves that honest miners do not choose to advocate for $SSM$ attack without been detected.

**$BWH$ Attacks.** Attackers can adopt $BWH$ attack to destroy rewards for the victim pool [20,21]. Attackers divide their mining power into innocent mining pool and infiltration mining pool. When infiltration pool finds a valid block (full proof of work, FPoW), he withholds it and continues to submit other shares (partial proof of work, PPoW) to obtain the share reward. The victim mining pool will never get rewards from the attacker's infiltration mining. Hence, the victim pool will suffer losses. Other miners, including innocent mining pool of adversary, will gain more rewards for the loss of the victim pool. [22] indicates that when attackers partition their mining power correctly, $BWH$ attack is more profitable than $HM$. However, when multiple independent pools adopt $BWH$ attack against each other (all pools have lower returns than $HM$), they will encounter the "miner's dilemma" [23].

**$FAW$ Attacks.** $FAW$ attack combines $SM$ and $BWH$ attacks [24]. In brief, $BWH$ attackers will discard the discovered FPoW, while in $FAW$, the attackers will reserve the FPoW. When other miners (not in the victim pool) find a valid blocks, the adversary will release and submit the previously reserved FPoW, causing a fork (similar to $SM$) to win in the forking competition and obtain share reward. In other cases, $FAW$ attack strategy is consistent with $BWH$. $FAW$ can get more rewards and avoid miner's dilemma compared with $BWH$. Attackers may succeed in forking competition, thereby obtaining the share reward. When attacker's branch is never selected as the main chain, $FAW$ will degenerate into $BWH$. Attackers with lower mining power will always fall into the miner's dilemma and lose profits when two attackers use $FAW$ attacks against each other, which is independent of their network environment. Conversely, attackers with higher mining power may avoid the miner's dilemma and gain higher profits, which is related to their network environment. [25] combines mining power adjustment strategies with $FAW$ attack ($PAW$), allowing attackers to adjust mining power dynamically between innocent mining and infiltration mining. Therefore, attackers can always increase their profits by allocating more mining power to more attractive mining strategies.

**Bribery Attacks.** Bribery attacks can increase the probability of the attacker's branch being selected as the main chain in forking competition [26]. Bribery attacks can only help the attacker win in the forking competition rather than bringing any profit to the attacker. Attackers can adopt origina3l bribery attack to win in forking competition, without obtaining any extra reward, instead. Therefore, original bribery attacks are always considered to combine with other attacks, such as double spending attack [27]. Bribery attack can be launched in a less visible way [28]. [25] combines bribery attack with $SM$. It indicates that compared with $SM$, bribery selfish mining ($BSM$) could bring 10% extra rewards to attackers. However, $BSM$ may cause the "venal miner's dilemma". [29] proposes an optimal $BSM$ to avoid the "venal miner's dilemma", where miners are considered perfectly rational. Attackers have lower mining power thresholds when making extra profits compared with $SM$. [30] proposes a mixed scenario where attackers alternate their strategy between $BWH$, $FAW$, and $PAW$. The mixed strategy is proved to be much higher in revenue than $HM$.

## 3 THREAT MODEL AND ASSUMPTION

### 3.1 Threat Model

An adversary can be an individual miner, or a mining pool formed by a collection of miners. Honest miners are profit-driven and could adopt the optimal mining strategy to increase their own profits without launching any mining attacks. Besides, adversaries can create different identities through sybil attacks and participate in multiple open mining pools with different accounts and IDs. Meanwhile, the adversary's mining power is limited to avoid 51% attack. He can allocate their mining power to innocent mining pool (similar to $HM$ strategy), selfish mining pool (similar to $SM$ strategy), or other mining attack strategies. More specifically, in $BSSM$ model, the adversary allocates their mining power to innocent mining pool and selfish mining pool. In $BSM$ model, adversaries only adopt $SM$. Finally, the adversary can create sybil nodes in the network to prioritize the propagation of their generated blocks, which increases the probability of selecting the attacker's branch as the main chain when forking occurs.

### 3.2 Assumption

To simplify our analysis, we make some reasonable assumptions. Our assumptions are similar to those of other selfish mining attacks, such as selfish mining [8], stubborn mining [16], semi-selfish mining [18] and bribery attacks [26, 34].

1. We normalized the total mining power of the system to 1. The (normalized) mining power of adversary is a value greater than 0 but less than 0.5, which is designed to avoid 51% attacks.

2. Miners are profit-driven. Honest miners can adopt the optimal mining strategy they consider to increase their profits, but will not launch mining attacks. This is reasonable because miners are honest but selfish. When the blockchain forks and the lengths of each branch are equal, miners could choose any branch.

3. There are no unintentional forks in the Bitcoin system. This assumption is rational because the probability of unintentional forks occurring in the Bitcoin system can be negligible, approximately 0.41% [31]. Therefore, combined with Assumption 1, the expected reward for a miner is equal to the probability of finding a valid block in each round. Due to the exponential distribution of the time for miners to find a valid block [32], average value is inversely proportional to their mining power, the probability of miners finding a valid block is equal to their normalized mining power.

4. We will normalize the coinbase reward for finding a valid block to 1 instead of 6.25 Bitcoins. In our analysis, miner's rewards are expected as well as normalized.

## 4 OBSERVATION AND MOTIVATION

### 4.1 Semi-selfish Mining

In semi-selfish mining, the adversary allocates mining power to the honest pools (similar to the honest mining strategy: mining as individual honest miners) and the selfish pools (similar to the selfish mining strategy: mining as selfish miners). In each round, the probability of honest pools generating a valid block is $\rho\alpha$, and the probability of selfish pools generating a valid block is $(1-\rho)\alpha$. Therefore, the probability of other pools generating a valid block is $1-\alpha$. The state transition process of semi-selfish mining is shown in Figure 1. The meanings of states $0, 0', 1, 2, 3, 4, ...$ are exactly the same as the states in selfish mining. On the basis, the states $1', 2', 3', 4', ...$ indicate that the last block in the public chain is generated by the adversary through honest pools, where the specific number represents the length of the private chain that the adversary reserves or hides.

Actually, there is a certain problem in analyzing the rewards of adversary while modeling semi-selfish mining,

which ignores the specific situations in which adversary may receive rewards. For example, when an attacker finds a valid block through honest pools, he will publish the block on the public chain and two blocks that are reserved (hidden) by selfish pools at once. The adversary will receive two block rewards regardless of which chain wins eventually (with probability $\alpha\rho$). In $BSSM$ reward analysis, we will revise this issue, as detailed in Section 6.2.

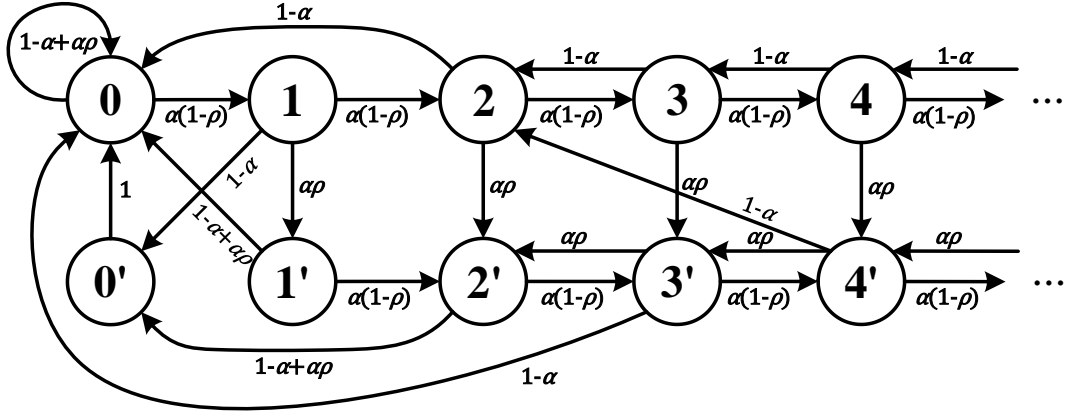

Figure 1: The state transition process of semi-selfish mining

## 4.2 Stubborn Mining

Stubborn mining extends the underlying model of selfish mining attacks. Its mining strategy is more "stubborn", which does not easily give up when leading, falling behind, and advancing together. In each round, the probability of selfish pools generating a valid block is $\alpha$, and the probability of other pools generating a valid block is $(1-\alpha)$. In addition, when the blockchain forks and the lengths of two branches are equal (one is a private chain of selfish pools, and the other is a public chain of other honest mining pools), the probability of other pools discovering a valid block and publishing it on the private chain of adversary is $\gamma(1-\alpha)$. Correspondingly, the probability of other pools publishing the block to public chain is $(1-\gamma)(1-\alpha)$.

Stubborn mining introduces three strategies by varying the degree of stubbornness of adversaries, which is designated as lead stubborn mining, equal-fork stubborn mining, and trail stubborn mining. The state transition process of three strategies of stubborn mining is shown in Figure 2. To simplify our analysis, we only discuss lead stubborn mining strategy. The meanings of states $0, 0', 1, 2, 3, ...$ are exactly the same as the states in selfish mining. The states $1', 2', 3', ...$ indicate that the blockchain forks and the lengths of two branches are equal (one is an adversary's private chain, and the other is a public chain of other honest mining pools), where the specific number indicates the length of hidden private chain of adversaries.

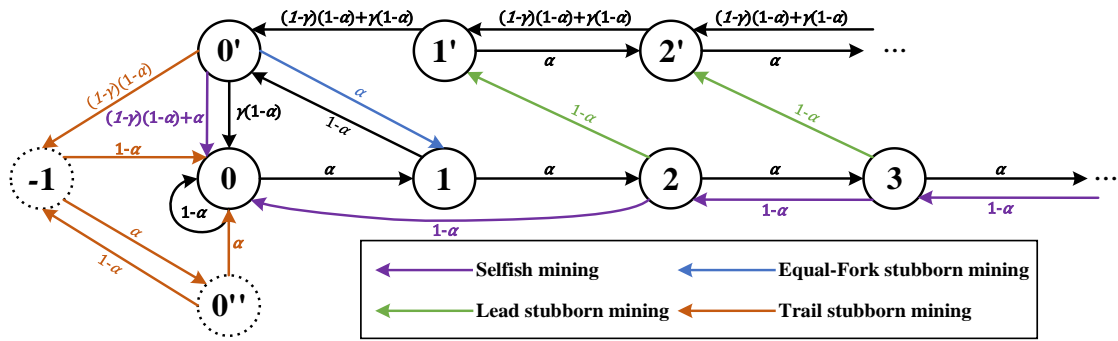

Figure 2: The state transition process of lead stubborn mining, equal-fork stubborn mining, and trail stubborn mining

## 4.3 Bribery Attack

When the blockchain forks and the lengths of two branches are equal, the adversaries may bribe some honest miners in other pools, which brings about the selfish branch of adversary a higher probability of successful competition thus becoming the main chain eventually. The process of bribery attack as shown in Figure 3. The part of bribed honest miners is called the target bribery pools. The reason why the target bribery pools are willing to accept bribes from the adversary is that the attackers will give a portion of the bribery money to the target bribery pools, which ensures that the total reward for the target bribery pools accepting bribes and expanding adversary's branches is no less than refusing bribes. Furthermore, the reward of adversary increases as the probability of the adversary's branch eventually becoming the main chain increases. When the adversaries choose to provide appropriate bribe money, they can obtain higher rewards than honest mining.

More specific, **(1)** when adversaries or target bribery pools find a valid block, they will publish it on private chain of adversary. Adversary's private chain wins and becomes the main chain with probability $(\alpha + \beta^b)$. **(2)** When other pools find a valid block, if they publish it on the public chain of other pools, other pools' public chain wins and becomes the main chain with probability $(1-\gamma)(1-\alpha-\beta^b)$. **(3)** If they publish it on the private chain of adversary, adversary's private chain wins and becomes the main chain with probability $\gamma(1-\alpha-\beta^b)$.

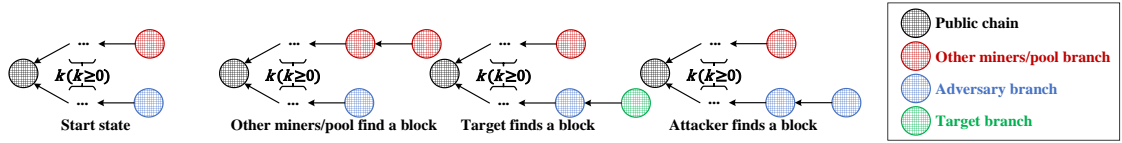

Figure 2: The process of bribery attack

## 5 BRIBERY SEMI-SELFISH MINING ($BSSM$)
### 5.1 Overview

We introduce bribery semi-selfish mining ($BSSM$) attack that combines bribery attack with semi-selfish mining. In the observation of bribery attack in Section 4.3, we point out that when the blockchain forks and the lengths of private branch of adversaries and public branch of other pools are equal, the adversaries may bribe some honest miners in other pools, increasing the probability of the private branch of adversary becoming the main chain. Therefore, $BSSM$ combines bribery attack with semi-selfish mining, which could increase the reward of adversary by adding bribery transactions on adversary's private branch.

Similar to $SSM$, adversary allocates mining power to the honest pools and selfish pools. We adopt $a$ to represent all adversary pools, $a_i$ to represent adversary's honest pools, and $a_s$ to indicate adversary's selfish pools. Accordingly, we use $b$ to represent target bribery pools, and $o$ to indicate other pools. When $a_s$ finds a valid block, he will reserve it. When another miner ($o$, $b$, or $a_i$) finds a valid block and publish it on public chain, adversaries will release a reserved block on the private chain at once, which brings about forking. $b$ will choose to mine on public branch (denying bribes) or mine on private branch of adversary (accepting bribes). Once $b$ chooses to expand private branch, he will claim to adversary that he accepts bribes. Otherwise, $b$ cannot claim to accept bribes from adversary. After the end of each round, adversary pays bribes to $b$ who accepts bribes.

### 5.2 Modeling $BSSM$

**State Transitions and probability.** We model the state transition process of $BSSM$ as shown in Figure 4. The meanings of states $k(k \geq 0)$ are exactly the same as the states in selfish mining. The states $k'(k \geq 1)$ indicate that the latest block on public chain is generated by $a_i$, and the private chain is reserved by $a_s$ before the block, where the number $k$ represents the difference between the length of the private chain and the public chain. More specifically, the length of the private chain reserved by $a_s$ is $(k+1)$. Note that the difference between states

$k'(k \geq 1)$ and states $k(k \geq 1)$ is that the former does not release the first reserved block on private chain. The reason is that the latest block on public chain in states $k'(k \geq 1)$ is generated by the adversary, while the latest block on public chain in states $k(k \geq 1)$ is generated by $o$. States $0'_0$, $0'_b$, and $0'_a$ represent the bribery initiation stage, where two branches of equal length appear in the system. In detail, state $0'_0$ indicates that two branches are formed by $a$ and $o$. State $0'_b$ represents that two branches are formed by $a$ and $b$. State $0'_a$ represents two branches are formed by $a_s$ and $a_i$. Next, we will discuss each state transition and probability in detail, as shown in Appendix A.

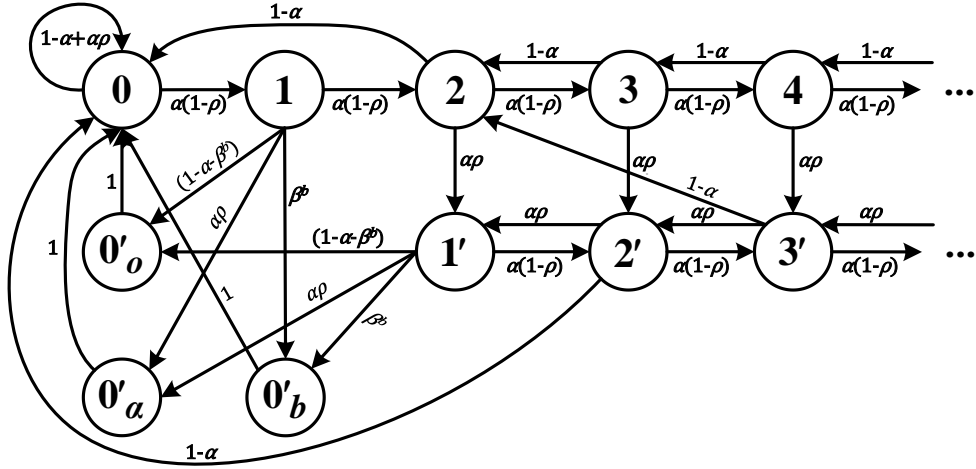

Figure 4: The state transition process of $BSSM$

According to Figure 4 of the state transition process of $BSSM$, we obtain the following equations:

$$\begin{cases} p_0 = (1 - \alpha + \rho\alpha)p_0 + (1 - \alpha)(p_2 + p_{2'}) + p_{0'_o} + p_{0'_b} + p_{0'_a} \\ p_1 = (1 - \rho)\alpha p_0 \\ p_{1'} = \rho\alpha(p_2 + p_{2'}) \\ p_{0'_o} = (1 - \alpha - \beta^b)(p_1 + p_{1'}) \\ p_{0'_b} = \beta^b(p_1 + p_{1'}) \\ p_{0'_a} = \rho\alpha(p_1 + p_{1'}) \\ p_k = (1 - \rho)\alpha p_{k-1} + (1 - \alpha)(p_{k+1} + p_{(k+1)'}), \text{when } k \geq 2 \\ p_{k'} = (1 - \rho)\alpha p_{(k-1)'} + \rho\alpha(p_{k+1} + p_{(k+1)'}), \text{when } k \geq 2 \\ \sum_{k=0}^{+\infty} p_k + \sum_{k=1}^{+\infty} p_{k'} + p_{0'_o} + p_{0'_b} + p_{0'_a} = 1 \end{cases} \quad (1)$$

**Reward.** We conduct a detailed analysis of the whole possible events (when a new block is generated). In $BSSM$, when adversaries have a certain block advantage through selfish mining, it does not mean that the adversary's private branch will win in the competition eventually, which is the most significant difference between $BSSM$ and $SM$. It is precisely for this reason that the difficulty of analyzing rewards for $a$, $b$, and $o$ has greatly increased. We observe from Figure 4 that states $k(k \geq 2)$ and states $k'(k \geq 2)$ will eventually transition to state 2 with probability $\frac{1-\alpha}{1-\alpha+\rho\alpha}$ or state $2'$ with probability $\frac{\rho\alpha}{1-\alpha+\rho\alpha}$. Therefore, based on states 2 and $2'$, we analyze the winning probability of private chain of $a$ and public chain of $o$ respectively in states $k(k \geq 2)$ or $k'(k \geq 2)$. Before analysis, we need to add two entities $P_b^p$ (represents the winning probability of public branch of $o$ in states $k(k \geq 2)$ or $k'(k \geq 2)$) and $P_b^s$ (represents the winning probability of private branch of $a$ in states $k(k \geq 2)$ or $k'(k \geq 2)$).

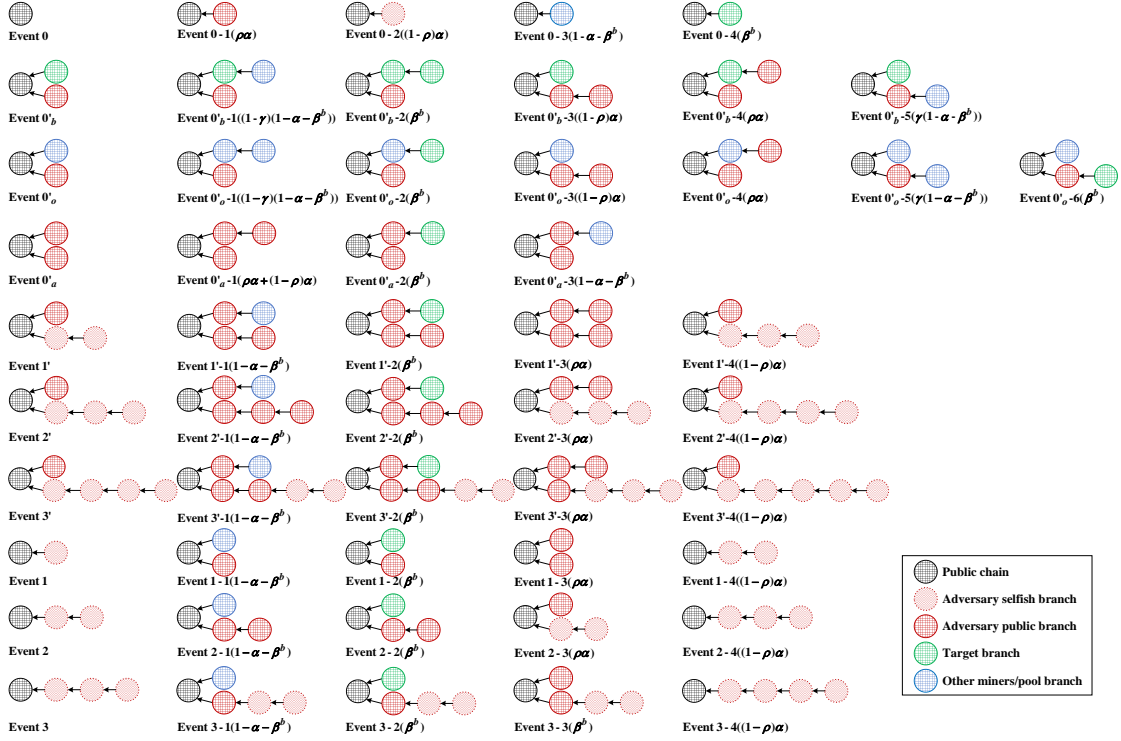

**Figure 5: Possible events in BSSM**

We observe event $0'_b$ in Figure 5: **(1)** when $o$ finds a valid block, he will publish it on public branch with probability $(1-\gamma)(1-\alpha-\beta^b)$ (public branch wins) or publish it on private branch with probability $\gamma(1-\alpha-\beta^b)$ (private branch wins); **(2)** when $b$ finds a valid block, he will publish it on public branch with probability $\beta^b$ (public branch wins); **(3)** when $a_s$ finds a valid block, he will publish it on private branch with probability $(1-\rho)\alpha$ (private branch wins); **(4)** when $a_i$ finds a valid block, he will publish it on public branch with probability $\rho\alpha$ (public branch wins). Similarly, we observe event $0'_0$: **(1)** when $o$ or $b$ finds a valid block, they will publish it on public branch with probability $((1-\gamma)(1-\alpha-\beta^b)+(1-\gamma)\beta^b)$ (public branch wins), or publish it on private branch with probability $(\gamma(1-\alpha-\beta^b)+\gamma\beta^b)$ (private branch wins); **(2)** when $a_s$ finds a valid block, he will publish it on private branch with probability $(1-\rho)\alpha$ (private branch wins); **(3)** when $a_i$ finds a valid block, he will publish it on public branch with probability $\rho\alpha$ (public branch wins). Finally, we observe event $0'_a$: **(1)** when $o$ or $b$ finds a valid block, they will publish it on public branch with probability $((1-\gamma)(1-\alpha-\beta^b)+(1-\gamma)\beta^b)$ (public branch wins), or publish it on private branch with probability $(\gamma(1-\alpha-\beta^b)+\gamma\beta^b)$ (private branch wins); **(2)** when $a_s$ finds a valid block, he will publish it on private branch with probability $(1-\rho)\alpha$ (private branch wins); **(3)** when $a_i$ finds a valid block, he will publish it on public branch with probability $\rho\alpha$ (public branch wins).

**Table 1: The state transitions of bribery initiation stage in BSSM**

| State $s$ | State $\hat{s}$ | $P_{0'_{\hat{s}}}$ |
|---|---|---|
| $k(k \geq 2)$ and $k'(k \geq 2)$ | $0'_a$ | $\dfrac{\rho\alpha}{1-\alpha+\rho\alpha} \cdot \dfrac{\rho\alpha}{1-\alpha-\beta^b+\beta^b+\rho\alpha}$ |
| $k(k \geq 2)$ and $k'(k \geq 2)$ | $0'_o$ | $\dfrac{\rho\alpha}{1-\alpha+\rho\alpha} \cdot \dfrac{1-\alpha-\beta^b}{1-\alpha-\beta^b+\beta^b+\rho\alpha}$ |
| $k(k \geq 2)$ and $k'(k \geq 2)$ | $0'_b$ | $\dfrac{\rho\alpha}{1-\alpha+\rho\alpha} \cdot \dfrac{\beta^b}{1-\alpha-\beta^b+\beta^b+\rho\alpha}$ |

Based on Figure 4, we can get the state transitions of bribery initiation stage in Table 1. Furthermore, we obtain the winning probability $P_b^s$ of private branch and $P_b^p$ of public branch in states $k(k \geq 2)$ and $k'(k \geq 2)$ as follows:

$$P_b^p = P_{0_b'}\left((1-\gamma)(1-\alpha-\beta^b) + \beta^b + \rho\alpha\right) + P_{0_o'}\left((1-\gamma)(1-\alpha-\beta^b) + (1-\gamma)\beta^b + \rho\alpha\right) \\ + P_{0_a'}\left((1-\gamma)(1-\alpha-\beta^b) + (1-\gamma)\beta^b + \rho\alpha\right) \quad (2)$$

$$P_b^s = \frac{1-\alpha}{1-\alpha+\rho\alpha} + P_{0_b'}(\gamma(1-\alpha-\beta^b) + (1-\rho)\alpha) + P_{0_o'}(\gamma(1-\alpha-\beta^b) + \gamma\beta^b + (1-\rho)\alpha) \\ + P_{0_a'}(\gamma(1-\alpha-\beta^b) + \gamma\beta^b + (1-\rho)\alpha) \quad (3)$$

Observing Figure 5, we continue to analyze the rewards of each event. For event 0: **(1)** when it transitions to event 0-1, $a$ gets 1 reward (probability $\rho\alpha$); **(2)** when it transitions to event 0-2, the rewards of $a$, $o$ and $b$ are determined later (probability $(1-\rho)\alpha$); **(3)** when it transitions to event 0-3, $o$ gets 1 reward (probability $(1-\alpha-\beta^b)$); **(4)** when it transitions to event 0-4, $b$ gets 1 reward (probability $\beta^b$). For event $0_b'$: **(1)** when it transitions to event $0_b'$-1, $o$ and $b$ get 1 reward (probability $(1-\gamma)(1-\alpha-\beta^b)$); **(2)** when it transitions to event $0_b'$-2, $b$ gets 2 rewards (probability $\beta^b$); **(3)** when it transitions to event $0_b'$-3, $a$ gets 2 rewards (probability $(1-\rho)\alpha$); **(4)** when it transitions to event $0_b'$-4, $a$ and $b$ get 1 reward (probability $\rho\alpha$); **(5)** when it transitions to event $0_b'$-5, $a$ and $o$ get 1 reward (probability $\gamma(1-\alpha-\beta^b)$). For event $0_o'$: **(1)** when it transitions to event $0_o'$-1, $o$ gets 2 rewards (probability $(1-\gamma)(1-\alpha-\beta^b)$); **(2)** when it transitions to event $0_o'$-2 and $b$ chooses to deny the bribes, $o$ and $b$ get 1 reward (probability $\beta^b$); **(3)** when it transitions to event $0_o'$-3, $a$ gets 2 rewards (probability $(1-\rho)\alpha$); **(4)** when it transitions to event $0_o'$-4, $a$ and $o$ get 1 reward (probability $\rho\alpha$); **(5)** when it transitions to event $0_o'$-5, $a$ and $o$ get 1 reward (probability $\gamma(1-\alpha-\beta^b)$); **(6)** when it transitions to event $0_o'$-6 and $b$ chooses to accept the bribes, $a$ and $b$ get 1 reward (probability $\beta^b$). For event $0_a'$: **(1)** when it transitions to event $0_a'$-1, $a$ gets 2 rewards (probability $\rho\alpha + (1-\rho)\alpha$); **(2)** when it transitions to event $0_a'$-2, $a$ and $b$ get 1 reward (probability $\beta^b$); **(3)** when it transitions to event $0_a'$-3, $a$ and $o$ get 1 reward (probability $(1-\alpha-\beta^b)$). For event $1'$: **(1)** when it transitions to event $1'$-1, $a$ gets 1 reward (probability $(1-\alpha-\beta^b)$); **(2)** when it transitions to event $1'$-2, $a$ gets 1 reward (probability $\beta^b$); **(3)** when it transitions to event $1'$-3, $a$ gets 1 reward (probability $\rho\alpha$); **(4)** when it transitions to event $1'$-4, the rewards of $a$, $o$ and $b$ are determined later (probability $(1-\rho)\alpha$). For event $2'$: **(1)** when it transitions to event $2'$-1, $a$ gets 3 rewards (probability $(1-\alpha-\beta^b)$); **(2)** when it transitions to event $2'$-2, $a$ gets 3 reward (probability $\beta^b$); **(3)** when it transitions to event $2'$-3, $a$ gets 1 reward (probability $\rho\alpha$); **(4)** when it transitions to Event $2'$-4, the rewards of $a$, $o$ and $b$ are determined later (probability $(1-\rho)\alpha$). For event $3'$: **(1)** when it transitions to event $3'$-1, $a$ gets $(1+P_b^s)$ rewards, $o$ gets $P_b^p$ reward (probability $(1-\alpha-\beta^b)$); **(2)** when it transitions to event $3'$-2, $a$ gets $(1+P_b^s)$ rewards, $b$ gets $P_b^p$ reward (probability $\beta^b$); **(3)** when it transitions to event $3'$-3, $a$ gets 1 reward (probability $\rho\alpha$); **(4)** when it transitions to event $2'$-4, the rewards of $a$, $o$ and $b$ are determined later (probability $(1-\rho)\alpha$). The reward analysis of events $k'(k > 3)$ is similar to event $3'$. For event 1: regardless of whether event 1 transitions to event 1-1 (probability $(1-\alpha-\beta^b)$), event 1-2 (probability $\beta^b$), event 1-3 (probability $\rho\alpha$), or event 1-4 (probability $(1-\rho)\alpha$), the rewards of $a$, $o$ and $b$ are determined later. For event 2: **(1)** when it transitions to event 2-1, $a$ gets 2 rewards (probability $(1-\alpha-\beta^b)$); **(2)** when it transitions to event 2-2, $a$ gets 2 rewards (probability $\beta^b$); **(3)** when it transitions to event 2-3 (probability $\rho\alpha$) or event 2-4 (probability $(1-\rho)\alpha$), the rewards of $a$, $o$ and $b$ are determined later. For event 3: **(1)** when it transitions to event 3-1, $a$ gets $P_b^s$ reward, $o$ gets $P_b^p$ reward (probability $(1-\alpha-\beta^b)$); **(2)** when it transitions to event 3-2, $a$ gets $P_b^s$ reward, $b$ gets $P_b^p$ reward (probability $\beta^b$); **(3)** when it transitions to event 3-3 (probability $\rho\alpha$) or event 3-4 (probability $(1-\rho)\alpha$), the rewards of $a$, $o$ and $b$ are determined later. The reward analysis of events $k(k > 3)$ is similar to event 3.

When $b$ chooses to accept the bribes, the $a$'s system reward $R_a$ is:

$$\begin{aligned}
R_a = &\; p_0 \cdot \rho\alpha + p_{0'_b} \cdot \left((1-\rho)\alpha \cdot 2 + \rho\alpha + \gamma(1-\alpha-\beta^b)\right) \\
&+ p_{0'_o} \cdot \left((1-\rho)\alpha \cdot 2 + \rho\alpha + \gamma(1-\alpha-\beta^b) + \beta^b\right) \\
&+ p_{0'_a} \cdot \left((\rho\alpha + (1-\rho)\alpha) \cdot 2 + \beta^b + (1-\alpha-\beta^b)\right) \\
&+ p_{1'} \cdot \left((1-\alpha-\beta^b) + \beta^b + \rho\alpha\right) + p_{2'} \cdot \left((1-\alpha-\beta^b) \cdot 3 + \beta^b \cdot 3 + \rho\alpha\right) \\
&+ \sum_{i=3}^{+\infty} p_{i'} \cdot \left((1-\alpha-\beta^b) \cdot (1+P_b^s) + \beta^b \cdot (1+P_b^s) + \rho\alpha\right) \\
&+ p_2 \cdot \left((1-\alpha-\beta^b) \cdot 2 + \beta^b \cdot 2\right) + \sum_{i=3}^{+\infty} p_i \cdot \left((1-\alpha-\beta^b) \cdot P_b^s + \beta^b \cdot P_b^s\right)
\end{aligned} \quad (4)$$

Obviously, $R_a$ is an increasing function with $\gamma$. That is to say, bribing more targets can bring more rewards to adversary.

When considering the bribes (a fraction $\varepsilon$ of the total system reward), the $a$'s reward $R_a^B$ is:
$$R_a^B = (1-\varepsilon)R_a \quad (5)$$

Obviously, $R_a^B$ is a decreasing function with $\varepsilon$. That is to say, paying more bribes to target can bring less rewards to adversary.

Accordingly, when $b$ chooses to accept the bribes and we consider the bribes, the $b$'s reward $R_b^B$ is:
$$\begin{aligned}
R_b^B = &\; p_0 \cdot \beta^b + p_{0'_b} \cdot \left((1-\gamma)(1-\alpha-\beta^b) + \beta^b \cdot 2 + \rho\alpha\right) + p_{0'_o} \cdot \beta^b + p_{0'_a} \cdot \beta^b \\
&+ \sum_{i=3}^{+\infty} p_{i'} \cdot \beta^b \cdot P_b^p + \sum_{i=3}^{+\infty} p_i \cdot \beta^b \cdot P_b^p + \varepsilon \cdot R_a
\end{aligned} \quad (6)$$

Obviously, $R_b^B$ is an increasing function with $\varepsilon$. That is to say, paying more bribes to targets can bring more rewards to $b$.

Finally, when $b$ chooses to accept the bribes and we consider the bribes, the $o$'s reward $R_o^B$ is:
$$\begin{aligned}
R_o^B = &\; p_0 \cdot (1-\alpha-\beta^b) + p_{0'_b} \cdot \left((1-\gamma)(1-\alpha-\beta^b) + \gamma(1-\alpha-\beta^b)\right) \\
&+ p_{0'_o} \cdot \left((1-\gamma)(1-\alpha-\beta^b) \cdot 2 + \rho\alpha + \gamma(1-\alpha-\beta^b)\right) + p_{0'_a} \cdot (1-\alpha-\beta^b) \\
&+ \sum_{i=3}^{+\infty} p_{i'} \cdot (1-\alpha-\beta^b) \cdot P_b^p + \sum_{i=3}^{+\infty} p_i \cdot (1-\alpha-\beta^b) \cdot P_b^p
\end{aligned} \quad (7)$$

Obviously, $R_o^B$ is a decreasing function with $\gamma$. That is to say, bribing more targets can bring less rewards to $o$.

Similarly, we consider when $b$ chooses to deny the bribes, the $a$'s reward $R_a^{B'}$ is:
$$\begin{aligned}
R_a^{B'} = &\; p_0 \cdot \rho\alpha + p_{0'_b} \cdot \left((1-\rho)\alpha \cdot 2 + \rho\alpha + \gamma(1-\alpha-\beta^b)\right) \\
&+ p_{0'_o} \cdot \left((1-\rho)\alpha \cdot 2 + \rho\alpha + \gamma(1-\alpha-\beta^b)\right) \\
&+ p_{0'_a} \cdot \left((\rho\alpha + (1-\rho)\alpha) \cdot 2 + \beta^b + (1-\alpha-\beta^b)\right) \\
&+ p_{1'} \cdot \left((1-\alpha-\beta^b) + \beta^b + \rho\alpha\right) + p_{2'} \cdot \left((1-\alpha-\beta^b) \cdot 3 + \beta^b \cdot 3 + \rho\alpha\right) \\
&+ \sum_{i=3}^{+\infty} p_{i'} \cdot \left((1-\alpha-\beta^b) \cdot (1+P_b^s) + \beta^b \cdot (1+P_b^s) + \rho\alpha\right) \\
&+ p_2 \cdot \left((1-\alpha-\beta^b) \cdot 2 + \beta^b \cdot 2\right) + \sum_{i=3}^{+\infty} p_i \cdot \left((1-\alpha-\beta^b) \cdot P_b^s + \beta^b \cdot P_b^s\right)
\end{aligned} \quad (8)$$

Accordingly, when $b$ chooses to deny the bribe, the $b$'s reward $R_b^{B'}$ is:
$$\begin{aligned}
R_b^{B'} = &\; p_0 \cdot \beta^b + p_{0'_b} \cdot \left((1-\gamma)(1-\alpha-\beta^b) + \beta^b \cdot 2 + \rho\alpha\right) \\
&+ p_{0'_o} \cdot \beta^b + p_{0'_a} \cdot \beta^b + \sum_{i=3}^{+\infty} p_{i'} \cdot \beta^b \cdot P_b^p + \sum_{i=3}^{+\infty} p_i \cdot \beta^b \cdot P_b^p
\end{aligned} \quad (9)$$

Finally, when $b$ chooses to deny the bribes, the $o$'s reward $R_o^{B'}$ is:
$$\begin{aligned}
R_o^{B'} = &\; p_0 \cdot (1-\alpha-\beta^b) + p_{0'_b} \cdot \left((1-\gamma)(1-\alpha-\beta^b) + \gamma(1-\alpha-\beta^b)\right) \\
&+ p_{0'_o} \cdot \left((1-\gamma)(1-\alpha-\beta^b) \cdot 2 + \beta^b + \rho\alpha + \gamma(1-\alpha-\beta^b)\right) \\
&+ p_{0'_a} \cdot (1-\alpha-\beta^b) + \sum_{i=3}^{+\infty} p_{i'} \cdot (1-\alpha-\beta^b) \cdot P_b^p + \sum_{i=3}^{+\infty} p_i \cdot (1-\alpha-\beta^b) \cdot P_b^p
\end{aligned} \quad (10)$$

THEOREM 5.1. *Once launching BSSM, the target $b$ can always obtain a higher reward when he chooses to accept the bribes at the bribery initiation stage.*

PROOF. Comparing the $b$'s reward $R_b^B$ when $b$ chooses to accept the bribes (Equation (6)) with the $b$'s reward $R_b^{B'}$ when $b$ chooses to deny the bribes (Equation (9)), we could derive $R_b^B \geq R_b^{B'}$ since $0 \leq \varepsilon \leq 1$ and $R_a > 0$. Once $a$ adopts $0 < \varepsilon \leq 1$, we could derive $R_b^B > R_b^{B'}$. Therefore, extending $a$'s private branch is always the optimal strategy at the bribery initiation stage.

THEOREM 5.2. *Once launching BSSM, $o$ is always forced to suffer losses when $b$ chooses to accept the bribes at the bribery initiation stage.*

PROOF. Comparing the $o$'s reward $R_o^B$ when $b$ chooses to accept the bribes (Equation (7)) with the $o$'s reward $R_o^{B'}$ when $b$ chooses to deny the bribes (Equation (10)), we could derive $R_o^B > R_o^{B'}$ since $\beta^b > 0$. Therefore, when $b$ chooses to accept the bribes at the bribery initiation stage, $o$ is always forced to suffer losses.

THEOREM 5.3. *Once launching BSSM, $a$ can obtain a higher reward than that in SSM when he pays proper bribes.*

PROOF. The rewards in $SSM$ are the same as the rewards in $BSSM$ when the target $b$ chooses to deny the bribes. Therefore, in order to obtain higher rewards, it is necessary for $a$ to ensure that $R_a^B > R_a^{B'}$. Comparing the $a$'s reward $R_a^B$ when $b$ chooses to accept the bribes (Equation (5)) with the $a$'s reward $R_a^{B'}$ when $b$ chooses to deny the bribes (Equation (8)), we could derive:

$$R_a^B > R_a^{B'} \Rightarrow \varepsilon < \frac{p_{o'_o} \cdot \beta^b}{p_{o'_o} \cdot \beta^b + R_a^{B'}} \quad (11)$$

The upper bound of $a$'s reward is $R_a$ in Equation (4) when $\varepsilon = 0$.

**Chain Growth Rate.** [8, 16, 18] indicates that the attack strategy based on selfish mining can lead to a decrease in the growth rate of the main chain. We note that the main chain here refers to the public chain generated by honest miners, rather than the private chain reserved by adversary. When adversary releases multiple reserved blocks, which makes the private chain longer than the public chain, and the private chain becomes the main chain eventually. According to the definition of the main chain growth rate, we calculate the main chain growth rate for $SM$, $SSM$, and $BSSM$ respectively:

$$\begin{cases} gr_{sm} = \alpha \cdot 0 + (1-\alpha) \cdot 1 \\ gr_{ssm} = (1-\rho)\alpha \cdot 0 + (1-\alpha) \cdot 1 + \rho\alpha \cdot 1 \\ gr_{bssm} = (1-\rho)\alpha \cdot 0 + (1-\alpha-\beta^b) \cdot 1 + \rho\alpha \cdot 1 + \beta^b \cdot 1 \end{cases} \quad (12)$$

THEOREM 5.4. *Once launching BSSM, the chain growth rate of BSSM and SSM are equal, and both are greater than the chain growth rate of SM.*

PROOF. Observing $gr_{sm}$, $gr_{ssm}$, and $gr_{bssm}$, we could derive $gr_{bssm} - gr_{sm} = gr_{bsm} - gr_{sm} = \rho\alpha$, which means $gr_{bssm} = gr_{ssm} > gr_{sm}$ since $\rho\alpha > 0$.

**Quantitative Analysis and Simulation.** Previous studies have shown that $SM$ can lead to a decrease in block generation rate (i.e., $R_a^B + R_o^B + R_b^B \leq 1$ and $R_a^{B'} + R_o^{B'} + R_b^{B'} \leq 1$). Therefore, we first normalize the relative reward entity $\tau(\frac{R_\tau^B}{R_a^B + R_o^B + R_b^B}, \frac{R_\tau^{B'}}{R_a^{B'} + R_o^{B'} + R_b^{B'}})$. Additionally, we use a specific example to demonstrate the adversary's relative extra reward when launching $BSSM$. Similar to [24], we adopt expected relative extra reward (RER) to evaluate $BSSM$. RER can be expressed as:

$$RER_\tau^{S_1, S_2} = \frac{R_\tau^{S_1} - R_\tau^{S_2}}{R_\tau^{S_2}} \quad (13)$$

$\tau$ represents an entity, which could be adversary ($a$), target bribery ($b$) pool or other pool ($o$). $S_1$ and $S_2$ represent different strategies, which include honest mining ($H$), semi-selfish mining ($SSM$), $b$ accepts the bribes in bribery semi-selfish mining ($BSSM$), $b$ denies the bribes in bribery semi-selfish mining ($BSSM'$), selfish mining ($SM$), $b$ accepts the bribes in bribery stubborn mining ($BSM$) and $b$ denies the bribes in bribery stubborn mining ($BSM'$). Therefore, $RER_\tau^{S_1}$ indicates the RER of entity $S_1$ when adopting mining strategy $\tau$. Obviously, $RER_a^H = \alpha$, which means the adversary's pool ($a$) who possesses mining power of $\alpha$ could obtain the RER of $\alpha$.

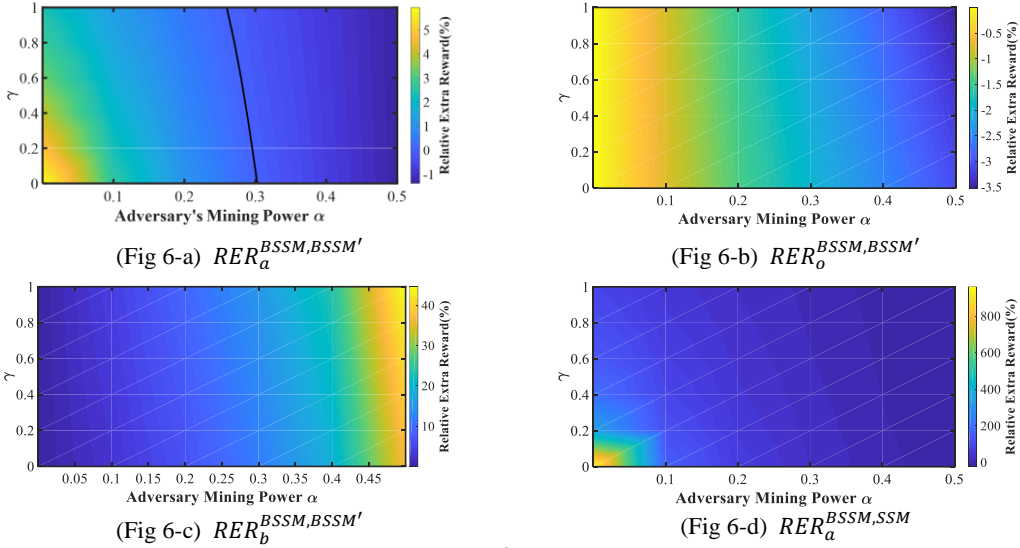

**Figure 6:** $RER$ when $\beta^b = 0.1, \varepsilon = 0.02, \rho = 0.1$.

First, we consider the RER of $a$, $o$ and $b$ in different strategies (accepting the bribes or denying the bribes) when $\beta^b = 0.1$, $\varepsilon = 0.02$ and $\rho = 0.1$. Fig 6-a, b, c shows the RER of $a$, $o$ and $b$ when accepting the bribes comparing with denying. As we expect, without considering $\gamma$, $a$ can obtain higher RER when he possesses less mining power. More specifically, the left side of the solid line indicates that $BSSM$ is the dominant strategy, while the right side of the solid line indicates that $BSSM'$ is the dominant strategy. Similar to our analysis results, the winning area of $BSSM$ as the dominant strategy is greater than that of $BSSM'$ as the dominant strategy. Based on THEOREM 5.3, $a$ can use a smaller $\varepsilon$ to expand the winning area in $BSSM$ (Fig 6-a). In addition, once launching $BSSM$, $o$ will always suffer losses (Fig 6-b). Fig 6-c shows that accepting bribes and expanding the private branch of adversary is always the optimal strategy for the target $b$ in bribery initiation state, which is consistent with THEOREM 5.1. Fig 6-d illustrates that no matter how much mining power the adversary possesses, $a$ prefers to launch $BSSM$ rather than adopt $SSM$. In detail, launching $BSSM$ will harm the profits of $o$, and is beneficial for $b$ to obtain higher RER. Adversaries with less mining power are more likely to get higher RER by launching $BSSM$, regardless of $\gamma$. This result indicates that the large mining pools lack sufficient motivation to launch $BSSM$.

Furthermore, we consider the RER of $a$ in different strategies ($BSSM$ or $BSSM'$) when $\rho = 0.1$, $\beta^b = 0.1$ or 0.3, comparing with $H$, $SM$ and $SSM$. We observe Figure 7, which indicates that adversary will definitely obtain higher rewards compared with denying the bribes when the target $b$ chooses to accept the bribes, regardless of $\alpha$ and $\gamma$. More specifically, Fig 7-a and b show the $RER_a^{BSSM,H}$ and $RER_a^{BSSM',H}$ when $\beta^b = 0.1$ or 0.3. A larger $\gamma$ will result in higher rewards for adversary, regardless of whether the target $b$ accepts or denies the bribes. Fig 7-c and d show the $RER_a^{BSSM,SM}$ and $RER_a^{BSSM',SM}$ when $\beta^b = 0.1$ or 0.3. Adversaries with small mining power could obtain higher rewards compared with $SM$ when launching $BSSM$ or $BSSM'$. In other words, the large mining pools have no motivation to launch $BSSM$ or $BSSM'$. Fig 7-e and f show the $RER_a^{BSSM,SSM}$ and $RER_a^{BSSM',SSM}$ when $\beta^b = 0.1$ or 0.3. Similarly, adversaries with small mining power are more profitable in launching $BSSM$ or $BSSM'$ compared with $SSM$. The RER of adversaries will decrease when $\alpha$ and $\gamma$ increase. The reason is that $\gamma$ represents the proportion of $o$ choosing to extend the private branch of adversaries.

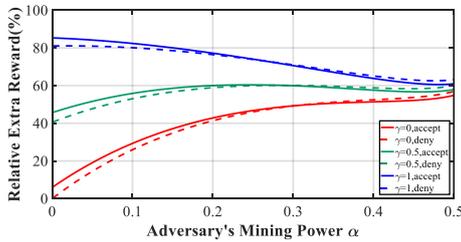

(Fig 7-a) $RER_a^{BSSM,H}$ and $RER_a^{BSSM',H}$ when $\beta^b = 0.1$

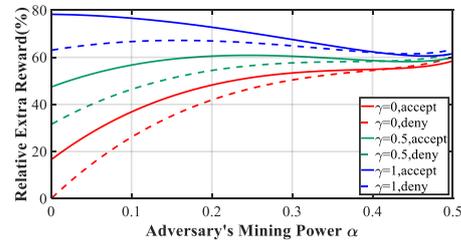

(Fig 7-b) $RER_a^{BSSM,H}$ and $RER_a^{BSSM',H}$ when $\beta^b = 0.3$

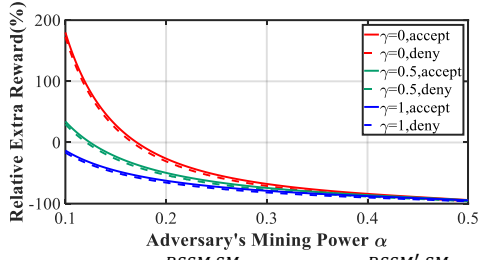

(Fig 7-c) $RER_a^{BSSM,SM}$ and $RER_a^{BSSM',SM}$ when $\beta^b = 0.1$

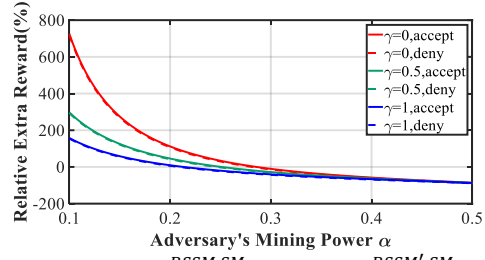

(Fig 7-d) $RER_a^{BSSM,SM}$ and $RER_a^{BSSM',SM}$ when $\beta^b = 0.3$

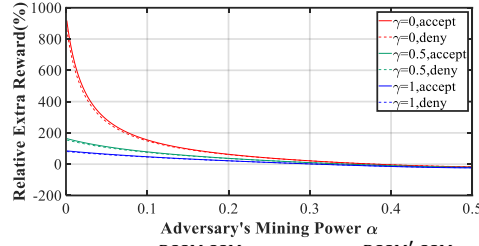

(Fig 7-e) $RER_a^{BSSM,SSM}$ and $RER_a^{BSSM',SSM}$ when $\beta^b = 0.1$

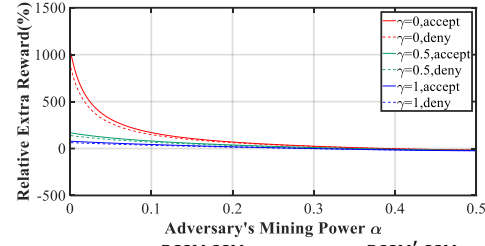

(Fig 7-f) $RER_a^{BSSM,SSM}$ and $RER_a^{BSSM',SSM}$ when $\beta^b = 0.3$

Figure 7: $RER_a$ when $\beta^b = 0.1$ or $\beta^b = 0.3$.

More specifically, we further consider the RER of $a$, $o$ and $b$ when the target $b$ chooses to accept the bribes compared with denying in Figure 8. The left side of the solid line in Fig 8-a indicates that adversaries can obtain higher rewards by adopting $BSSM$ compared with $BSSM'$. Conversely, the right side of the solid line represents that $BSSM'$ is the optimal strategy for adversary. Besides, the RER of adversaries will increase when $\rho$ (the proportion of adversary adopting honest mining) decreases. Fig 8-b indicates that once the target $b$ chooses to accept the bribes, $o$ will suffer losses. Similarly, Fig 8-c shows that the target $b$ prefers accepting the bribes to denying, which means choosing to accept the bribes is always the optimal strategy. The simulation results are completely consistent with our previous theoretical analysis of the RER of $a$, $o$ and $b$.

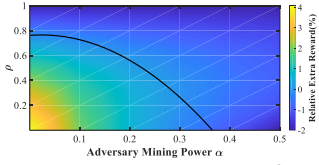

(Fig 8-a) $RER_a^{BSSM,BSSM'}$

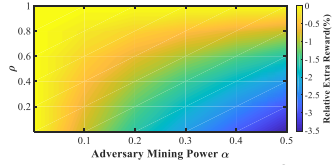

(Fig 8-b) $RER_o^{BSSM,BSSM'}$

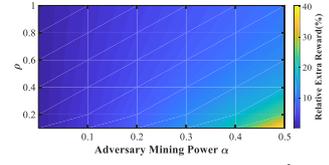

(Fig 8-c) $RER_b^{BSSM,BSSM'}$

Figure 8: $RER_{a,o,b}^{BSSM,BSSM'}$ when $\beta^b = 0.1, \varepsilon = 0.02, \gamma = 0.5$.

Finally, we consider the chain growth rate when adversaries adopt different strategies ($SM$ or $BSSM$). More specifically, Figure 9 shows $BSSM$ and $SM$'s chain growth rate when $\rho = 0.1$ or $0.3$. As expected, the higher mining power adversaries possess, the smaller the chain growth rate. This is because there is an inverse correlation between the mining power of $o$ and $a$. The growth rate of the main chain mainly depends on $o$'s ability to discover a new block. In addition, a larger $\rho$ will increase the probability of adversary generating a new block in the main chain. The above simulation results are consistent with our previous theoretical analysis of chain growth rate.

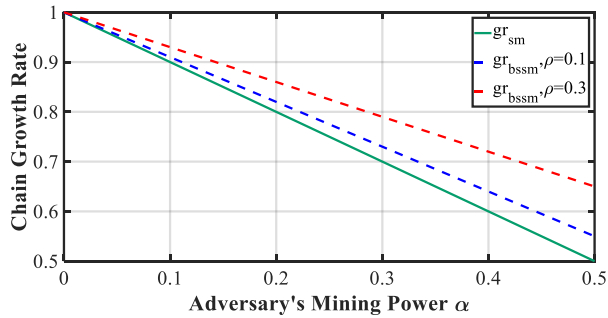

Figure 9: $BSSM$ and $SM$'s chain growth rate

## 6 BRIBERY STUBBORN MINING ($BSM$)
### 6.1 Overview
We introduce bribery stubborn mining ($BSM$) attack that combines bribery attack with stubborn mining, which could increase the reward of adversary by adding bribery transactions on adversary's private branch. In $BSM$, the adversaries adopt selfish mining with the whole mining power. We adopt $a$ to represent all adversary pools. Once $a$ finds a valid block, he will reserve it and form private chain. However, when another miner ($o$ or $b$) finds a valid block, he will publish it on the public chain, and then $a$ will release a reserved block at once, which brings about forking. $b$ will choose to mine on public branch (denying bribes) or mine on private branch of adversary (accepting bribes). The bribery payment process is similar to $BSSM$.

### 6.2 Modeling $BSM$
**State Transitions and probability.** We model the state transition process of $BSM$ as shown in Figure 10. The meanings of states $k(k \geq 0)$ and states $k'(k \geq 1)$ are exactly the same as the states in stubborn mining. States $0'_o$ and $0'_b$ represent the bribery initiation stage. More specifically, state $0'_o$ indicates that two branches are formed by $a$ and $o$. State $0'_b$ represents that two branches are formed by $a$ and $b$. Next, we will discuss each state transition and probability in detail, as shown in Appendix B.

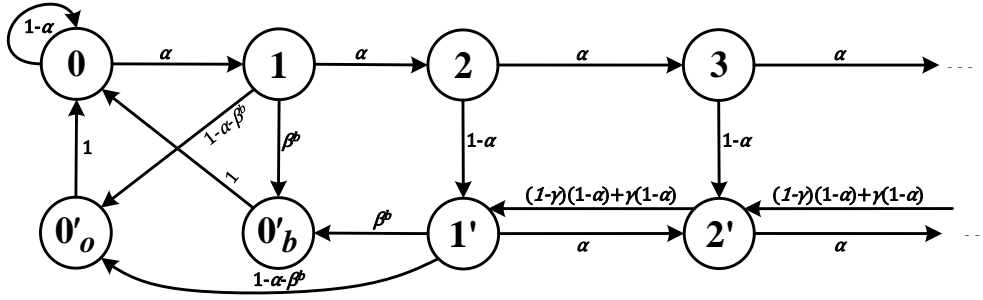

Figure 10: The state transition process of $BSM$

According to Figure 10 of the state transition process of $BSM$, we obtain the following equations:

$$\begin{cases} p_0 = (1-\alpha)p_0 + p_{0'_o} + p_{0'_b} \\ p_{0'_o} = (1-\alpha-\beta^b)(p_1 + p_{1'}) \\ p_{0'_b} = \beta^b(p_1 + p_{1'}) \\ p_{1'} = (1-\alpha)(p_2 + p_{2'}) \\ p_k = \alpha p_{k-1}, \text{when } k \geq 1 \\ p_{k'} = \alpha p_{(k-1)'} + (1-\alpha)(p_{k+1} + p_{(k+1)'}), \text{when } k \geq 2 \\ \sum_{k=0}^{+\infty} p_k + \sum_{k=1}^{+\infty} p_{k'} + p_{0'_o} + p_{0'_b} = 1 \end{cases} \quad (13)$$

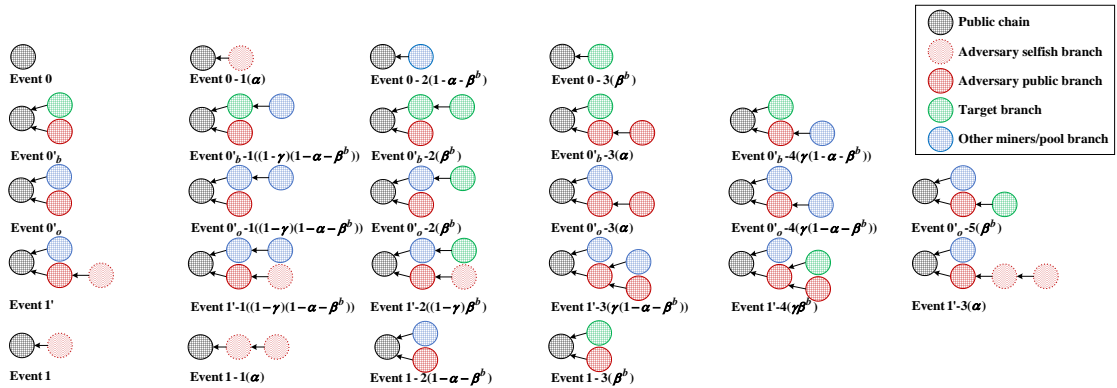

Figure 11: Possible events in $BSM$

**Reward.** We conduct a detailed analysis of the whole possible events. We observe from Figure 10 that they will eventually transition to state $0'_0$ with probability $(1-\alpha-\beta^b)$ or state $0'_b$ with probability $\beta^b$ whether how many block advantages the adversaries possess through selfish mining. Therefore, we need to analyze the winning probability of private chain of $a$ and public chain of $o$ respectively. Before analysis, we need to add two entities $P_b^p$ (represents the winning probability of public branch of $o$ and $P_b^s$ (represents the winning probability of private branch of $a$).

We observe event $0'_b$ in Figure 11: **(1)** when $o$ finds a valid block, he will publish it on public branch with probability $(1-\gamma)(1-\alpha-\beta^b)$ (public branch wins) or publish it on private branch with probability $\gamma(1-\alpha-\beta^b)$ (private branch wins); **(2)** when $b$ finds a valid block, he will publish it on public branch with probability $\beta^b$ (public branch wins); **(3)** when $a$ finds a valid block, he will publish it on private branch with probability $\alpha$ (private branch wins). Similarly, we observe event $0'_0$: **(1)** when $o$ or $b$ finds a valid block, they will publish it on public branch with probability $((1-\gamma)(1-\alpha-\beta^b)+(1-\gamma)\beta^b)$ (public branch wins), or publish it on private branch with probability $(\gamma(1-\alpha-\beta^b)+\gamma\beta^b)$ (private branch wins); **(2)** when $a$ finds a valid block, he will publish it on private branch with probability $\alpha$ (private branch wins).

For states $k(k\geq 1)$ and states $k'(k\geq 1)$, they transition to state $0'_b$ with probability $P_{0'_b}=\frac{\beta^b}{1-\alpha-\beta^b+\beta^b}$, and transition to the state $0'_0$ with probability $P_{0'_o}=\frac{1-\alpha-\beta^b}{1-\alpha-\beta^b+\beta^b}$. We further derive the winning probability $P_b^s$ of private branch and $P_b^p$ of public branch in states $k(k\geq 1)$ and $k'(k\geq 1)$ as follows:

$$\begin{cases} P_b^p = P_{0'_b}\left((1-\gamma)(1-\alpha-\beta^b)+\beta^b\right) + P_{0'_o}\left((1-\gamma)(1-\alpha-\beta^b)+(1-\gamma)\beta^b\right) \\ P_b^s = P_{0'_b}(\gamma(1-\alpha-\beta^b)+\alpha) + P_{0'_o}(\gamma(1-\alpha-\beta^b)+\gamma\beta^b+\alpha) \end{cases} \quad (14)$$

Observing Figure 11, we continue to analyze the rewards of each event. For event 0: **(1)** when it transitions to event 0-1, the rewards of $a$, $o$ and $b$ are determined later (probability $\alpha$); **(2)** when it transitions to event 0-2, $o$ get 1 reward (probability $(1-\alpha-\beta^b)$); **(3)** when it transitions to event 0-3, $b$ gets 1 reward (probability $\beta$). For event $0'_b$: **(1)** when it transitions to event $0'_b$-1, $o$ and $b$ get 1 reward (probability $(1-\gamma)(1-\alpha-\beta^b)$); **(2)** when it transitions to event $0'_b$-2, $b$ gets 2 rewards (probability $\beta^b$); **(3)** when it transitions to event $0'_b$-3, $a$ gets 2 rewards (probability $\alpha$); **(4)** when it transitions to event $0'_b$-4, $a$ and $o$ get 1 reward (probability $\gamma(1-\alpha-\beta^b)$). For event $0'_o$: **(1)** when it transitions to event $0'_o$-1, $o$ gets 2 rewards (probability $(1-\gamma)(1-\alpha-\beta^b)$); **(2)** when it transitions to event $0'_o$-2 and $b$ chooses to deny the bribes, $o$ and $b$ get 1 reward (probability $\beta^b$); **(3)** when it transitions to event $0'_o$-3, $a$ gets 2 rewards (probability $\alpha$); **(4)** when it transitions to event $0'_o$-4, $a$ and $o$ get 1 reward (probability $\gamma(1-\alpha-\beta^b)$); **(5)** when it transitions to event $0'_o$-5 and $b$ chooses to accept the bribes, $a$ and $b$ get 1 reward (probability $\beta^b$). For event $1'$: **(1)** when it transitions to event $1'$-1, $a$ gets $P_b^s$ reward, $o$ gets $P_b^p$ reward (probability $(1-\gamma)(1-\alpha-\beta^b)$); **(2)** when it transitions to event $1'$-2, $a$ gets $P_b^s$ reward, $o$ get $P_b^p$ reward (probability $\beta^b$); **(3)** when it transitions to event $1'$-3, $a$ get 1 reward (probability $\gamma(1-\alpha-\beta^b)$); **(4)** when it transitions to event $1'$-4, $a$ get 1 reward (probability $\beta^b$); **(5)** when it transitions to event $1'$-5, the rewards of $a$, $o$ and $b$ are determined later (probability $\alpha$). The reward analysis of events $k'(k>2)$ is similar to event $1'$. For event 1: the rewards of $a$, $o$ and $b$ are determined later whether it transitions to event 1-1 (probability $\alpha$), event 1-2 (probability $(1-\alpha-\beta^b)$) or event 1-3 (probability $\beta^b$). The reward analysis of events $k(k\geq 2)$ is similar to event 1.

When $b$ chooses to accept the bribes, the $a$'s system reward $R_a$ is:

$$\begin{aligned} R_a &= p_{0'_b}\cdot(\alpha\cdot 2+\gamma(1-\alpha-\beta^b)) + p_{0'_o}\cdot(\alpha\cdot 2+\gamma(1-\alpha-\beta^b)+\beta^b) \\ &\quad + \sum_{i=1}^{+\infty}p_{i'}\cdot\left((1-\gamma)(1-\alpha-\beta^b)\cdot P_b^s + \gamma(1-\alpha-\beta^b) + (1-\gamma)\beta^b\cdot P_b^s + \gamma\beta^b\right) \\ &= p_{0'_b}\cdot(\alpha\cdot 2+\gamma(1-\alpha-\beta^b)) + p_{0'_o}\cdot(\alpha\cdot 2+\gamma(1-\alpha-\beta^b)+\beta^b) \\ &\quad + \sum_{i=1}^{+\infty}p_{i'}\cdot\left(\gamma(1-P_b^s)+P_b^s\right)(1-\alpha) \end{aligned} \quad (15)$$

When considering the bribes (a fraction $\varepsilon$ of the total system reward), the $a$'s reward $R_a^B$ is:
$$R_a^B = (1-\varepsilon)R_a \tag{16}$$

Accordingly, when $b$ chooses to accept the bribes and we consider the bribes, the $b$'s reward $R_b^B$ is:
$$R_b^B = p_0 \cdot \beta^b + p_{0'_b} \cdot \left((1-\gamma)(1-\alpha-\beta^b) + \beta^b \cdot 2\right) + p_{0'_o} \cdot \beta^b + \varepsilon \cdot R_a \tag{17}$$

Finally, when $b$ chooses to accept the bribes and we consider the bribes, the $o$'s reward $R_o^B$ is:
$$\begin{aligned}R_o^B &= p_0 \cdot (1-\alpha-\beta^b) + p_{0'_b} \cdot \left((1-\gamma)(1-\alpha-\beta^b) + \gamma(1-\alpha-\beta^b)\right) \\ &\quad + p_{0'_o} \cdot \left((1-\gamma)(1-\alpha-\beta^b) \cdot 2 + \gamma(1-\alpha-\beta^b)\right) \\ &\quad + \sum_{i=1}^{+\infty} p_{i'} \cdot \left((1-\gamma)(1-\alpha-\beta^b) \cdot P_b^p + (1-\gamma)\beta^b \cdot P_b^p\right) \\ &= p_0 \cdot (1-\alpha-\beta^b) + p_{0'_b} \cdot (1-\alpha-\beta^b) + p_{0'_o} \cdot (2-\gamma)(1-\alpha-\beta^b) \\ &\quad + \sum_{i=1}^{+\infty} p_{i'} \cdot \left((1-\gamma)(1-\alpha-\beta^b) \cdot P_b^p + (1-\gamma)\beta^b \cdot P_b^p\right)\end{aligned} \tag{18}$$

Similarly, we consider when $b$ chooses to deny the bribes, the $a$'s reward $R_a^{B'}$ is:
$$\begin{aligned}R_a^{B'} &= p_{0'_b} \cdot (\alpha \cdot 2 + \gamma(1-\alpha-\beta^b)) + p_{0'_o} \cdot \left(\alpha \cdot 2 + \gamma(1-\alpha-\beta^b)\right) \\ &\quad + \sum_{i=1}^{+\infty} p_{i'} \cdot \left((1-\gamma)(1-\alpha-\beta^b) \cdot P_b^s + \gamma(1-\alpha-\beta^b) + (1-\gamma)\beta^b \cdot P_b^s + \gamma\beta^b\right)\end{aligned} \tag{19}$$

Accordingly, when $b$ chooses to deny the bribe, the $b$'s reward $R_b^{B'}$ is:
$$R_b^{B'} = p_0 \cdot \beta^b + p_{0'_b} \cdot \left((1-\gamma)(1-\alpha-\beta^b) + \beta^b \cdot 2\right) + p_{0'_o} \cdot \beta^b \tag{20}$$

Finally, when $b$ chooses to deny the bribes, the $o$'s reward $R_o^{B'}$ is:
$$\begin{aligned}R_o^{B'} &= p_0 \cdot (1-\alpha-\beta^b) + p_{0'_b} \cdot \left((1-\gamma)(1-\alpha-\beta^b) + \gamma(1-\alpha-\beta^b)\right) \\ &\quad + p_{0'_o} \cdot \left((1-\gamma)(1-\alpha-\beta^b) \cdot 2 + \beta^b + \gamma(1-\alpha-\beta^b)\right) \\ &\quad + \sum_{i=1}^{+\infty} p_{i'} \cdot \left((1-\gamma)(1-\alpha-\beta^b) \cdot P_b^p + (1-\gamma)\beta^b \cdot P_b^p\right)\end{aligned} \tag{21}$$

THEOREM 6.1. *Once launching BSM, the target $b$ can always obtain a higher reward when he chooses to accept the bribes at the bribery initiation stage.*

PROOF. Comparing the $b$'s reward $R_b^B$ when $b$ chooses to accept the bribes (Equation (17)) with the $b$'s reward $R_b^{B'}$ when $b$ chooses to deny the bribes (Equation (20)), we could derive $R_b^B \geq R_b^{B'}$ since $0 \leq \varepsilon \leq 1$ and $R_a > 0$. Once $a$ adopts $0 < \varepsilon \leq 1$, we could derive $R_b^B > R_b^{B'}$. Therefore, extending $a$'s private branch is always the optimal strategy at the bribery initiation stage.

THEOREM 6.2. *Once launching BSM, $o$ is always forced to suffer losses when $b$ chooses to accept the bribes at the bribery initiation stage.*

PROOF. Comparing the $o$'s reward $R_o^B$ when $b$ chooses to accept the bribes (Equation (18)) with the $o$'s reward $R_o^{B'}$ when $b$ chooses to deny the bribes (Equation (21)), we could derive $R_o^B > R_o^{B'}$ since $\beta^b > 0$. Therefore, when $b$ chooses to accept the bribes at the bribery initiation stage, $o$ is always forced to suffer losses.

THEOREM 6.3. *Once launching BSM, $a$ can obtain a higher reward than that in stubborn mining when he pays proper bribes.*

PROOF. The rewards in stubborn mining are the same as the rewards in $BSM$ when the target $b$ chooses to deny the bribes. Therefore, in order to obtain higher rewards, it is necessary for $a$ to ensure that $R_a^B > R_a^{B'}$. Comparing the $a$'s reward $R_a^B$ when $b$ chooses to accept the bribes (Equation (16)) with the $a$'s reward $R_a^{B'}$ when $b$ chooses to deny the bribes (Equation (19)), we could derive:

$$R_a^B > R_a^{B'} \Rightarrow \varepsilon < \frac{p_{0'_o} \cdot \beta^b}{p_{0'_o} \cdot \beta^b + R_a^{B'}} \tag{22}$$

The upper bound of $a$'s reward is $R_a$ in Equation (15) when $\varepsilon = 0$.

**Quantitative Analysis and Simulation.** We use the RER in Equation (13) to evaluate $BSM$. First, we consider the RER of $a$, $o$ and $b$ in different strategies (accepting the bribes or denying the bribes) when $\beta^b = 0.1$, $\varepsilon = 0.02$. Fig 12-a, b, c shows the RER of $a$, $o$ and $b$ when accepting the bribes comparing with denying. As we expect, without considering $\gamma$, $a$ can obtain higher RER when he possesses less mining power. More specifically, the left side of the solid line indicates that $BSM$ is the dominant strategy, while the right side of the solid line indicates that $BSM'$ is the dominant strategy. Obviously, the winning area of $BSM$ as the dominant strategy is greater than that of $BSM'$. Based on THEOREM 6.3, $a$ can use a smaller $\varepsilon$ to expand the winning area in $BSM$ (Fig 12-a). In addition, once launching $BSM$, $o$ will always suffer losses (Fig 12-b). Fig 12-c shows that accepting bribes and expanding the private branch of adversary is always the optimal strategy for the target $b$ in bribery initiation state, which is consistent with THEOREM 6.1. In detail, launching $BSM$ will harm the profits of $o$, and is beneficial for $b$ to obtain higher RER. Adversaries with less mining power are more likely to get higher RER by launching $BSM$, regardless of $\gamma$. This result indicates that the large mining pools lack sufficient motivation to launch $BSM$.

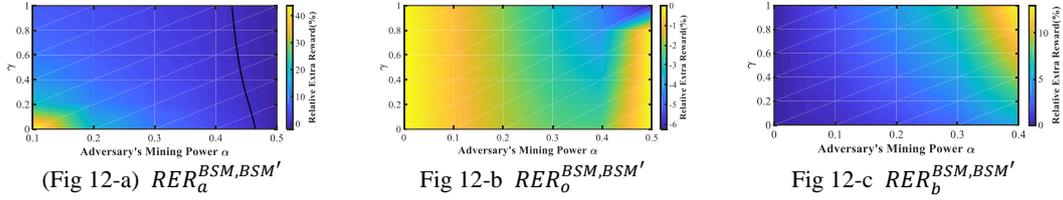

(Fig 12-a) $RER_a^{BSM,BSM'}$  Fig 12-b $RER_o^{BSM,BSM'}$  Fig 12-c $RER_b^{BSM,BSM'}$

**Figure 12: $RER_{a,o,b}^{BSM,BSM'}$ when $\boldsymbol{\beta^b = 0.1, \varepsilon = 0.02}$.**

Furthermore, we consider the RER of $a$ in $BSM$ when $\beta^b = 0.1$, $\varepsilon = 0.02$, comparing with $H$ and $SM$ in Figure 13. We observe Fig 13-a, which shows that adversaries have an advantage in adopting $BSM$ compared with $SM$ in specific situations. More specifically, the upper side of the solid line indicates that adopting $BSM$ is more profitable for adversaries, while the lower side of the solid line shows that launching $SM$ is the optimal strategy. Fig 13-b depicts the RER of adversaries when $\beta^b = 0.1$, $\varepsilon = 0.02$, comparing with $H$. Furthermore, the left side of the solid line indicates that $H$ is the optimal strategy, while the right side of the solid line shows that adopting $BSM$ is more profitable for adversaries. As expected, adversaries with high mining power have more motivation to launch $BSM$.

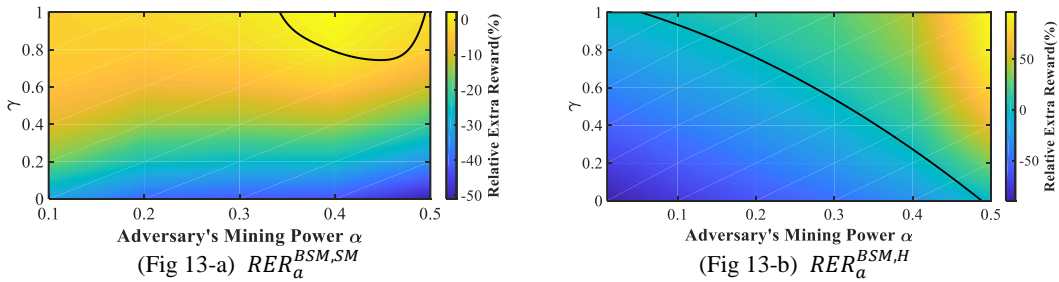

(Fig 13-a) $RER_a^{BSM,SM}$  (Fig 13-b) $RER_a^{BSM,H}$

**Figure 13: $RER_a$ when $\boldsymbol{\beta^b = 0.1, \varepsilon = 0.02}$.**

Finally, we consider the RER of adversaries in different strategies ($BSM$ and $BSM'$) when $\beta^b = 0.1$ or $\beta^b = 0.3$, comparing with $H$ in Figure 14. Figure 14 shows that choosing to accept the bribes is always the optimal strategy for adversaries. More specifically, Fig 14-a shows the $RER_a^{BSM,H}$ and $RER_a^{BSM',H}$ when $\beta^b = 0.1$, and Fig 14-b shows the $RER_a^{BSM,H}$ and $RER_a^{BSM',H}$ when $\beta^b = 0.3$. A larger $\gamma$ will result in higher rewards for adversary, regardless of whether the target $b$ accepts or denies the bribes, which is consistent with our theoretical analysis of $BSM$.

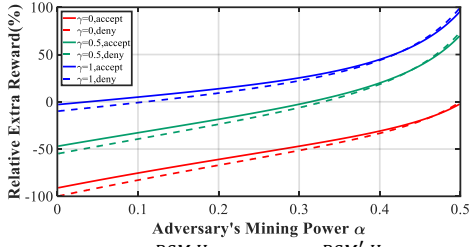 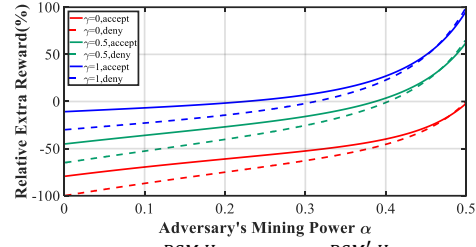

(Fig 14-a) $RER_a^{BSM,H}$ and $RER_a^{BSM',H}$ when $\beta^b = 0.1$  (Fig 14-b) $RER_a^{BSM,H}$ and $RER_a^{BSM',H}$ when $\beta^b = 0.3$

Figure 14: $RER_a$ when $\beta^b = 0.1$ or $\beta^b = 0.3$.

## 7 The Bribery Miner's Dilemma

In selfish mining and $BSM$, adversary can get extra rewards by deliberately forking. Previous work has pointed out that the adversary's extra reward comes from the loss of $b$ and $o$ [8,24,25]. More detail, once $o$ expands the private branch of $a$ instead of the public branch of $b$ (event $0'_b$-3 in $BSSM$ and $BSM$), $b$ will suffer losses. Nevertheless, $b$ cannot avoid losses regardless of the strategy adopted by the target $b$. More specifically, whether target b suffers losses is controlled by the strategy of $o$, rather than by themselves. Meanwhile, when b chooses to accept the bribes, $o$ will suffer losses (event $0'_o$-6 in $BSSM$ and event $0'_o$-5 in $BSM$).

We have demonstrated that the optimal strategy for target $b$ in $BSSM$ and $BSM$ is to accept the bribes and extend the adversary's private branch. Therefore, adversary can bribe multiple targets simultaneously, which increases the winning probability of private branch of $a$. In this case, multiple targets who accept the bribes may fall into the "bribery miner's dilemma": all targets $b$s will suffer losses due to accepting the bribes (similar to the "miner's dilemma"). When the whole targets deny the bribes, they will obtain extra reward, comparing with adopting honest mining. But for each target $b$, they would not choose to deny the bribes. This is because accepting the bribes is always a locally optimal strategy for $b$ at bribery initiation stage. Therefore, we have a single Nash equilibrium for targets under $BSSM$ and $BSM$: all targets will choose to accept the bribes and expand adversary's branch at the bribery initiation stage.

### 7.1 The "Bribery Miner's Dilemma" in $BSSM$

In $BSSM$, we consider two targets $b_1$ and $b_2$ with mining power $\beta_1^b$ and $\beta_2^b$. We set $\alpha = 0.3$, $\beta_1^b = 0.2$, $\rho = 0.1$ and $\varepsilon = 0.02$. We define target $b_i$'s winning condition is to obtain a higher reward than honest mining (i.e., $RER_{b_i}^{BSSM,H} > 0$). We calculate the rewards of $b_1$, $b_2$, $a$ and $o$ respectively in four cases ((**1**) both $b_1$ and $b_2$ accept the bribes; (**2**) $b_1$ accepts the bribes but $b_2$ denies; (**3**) $b_2$ accepts the bribes but $b_1$ denies; (**4**) both $b_1$ and $b_2$ deny the bribes). Figure 15 shows the RER and winning conditions for each target in terms of $\beta_2^b$ and $\gamma$.

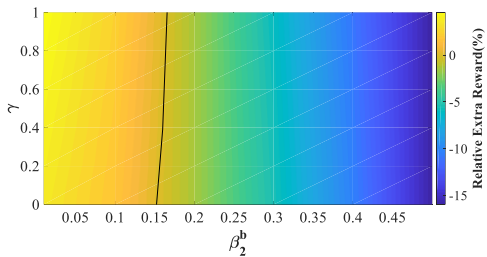 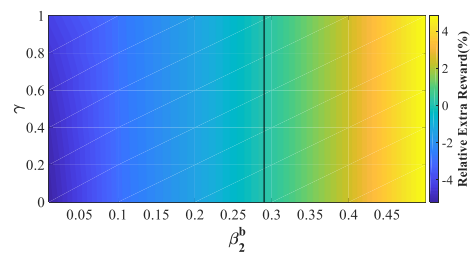

(Fig 15-a) $RER_{b_1}^{BSSM,H}$ and winning condition    (Fig 15-b) $RER_{b_2}^{BSSM,H}$ and winning condition

Figure 15: RER and winning conditions in $BSSM$

The left side of the solid line in Fig 15-a represents the winning condition of target $b_1$, and the right side of solid line in Fig 15-b represents the winning condition of target $b_2$. Fig 15-a indicates that target $b_1$ will obtain extra reward while $b_2$ will suffer losses when $\beta_2^b$ is relatively small. Fig 15-b indicates that target $b_2$ will obtain extra reward while $b_1$ will suffer losses when $\beta_2^b$ is relatively large. The area between two solid lines represents both $b_1$ and $b_2$ will suffer losses (i.e., they will encounter the "bribery miner's dilemma"). The RER of target $b_1$ and $b_2$ will not be greatly affected by $\gamma$ whether $\beta_2^b$ is large or small. This is because the change of $\gamma$ will bring $RER_{b_i}^{BSSM,H}$ to the same trend of change. For the adversary, when proper value of $\beta_1^b$, $\beta_2^b$, $\rho$ and $\varepsilon$, the adversary could obtain extra reward and make the target $b_1$ and $b_2$ fall into the "bribery miner's dilemma".

We use a more intuitive example to demonstrate the "briery miner's dilemma" in $BSSM$. We set $\gamma = 0$, $\varepsilon = 0.02$ and $\rho = 0.1$, and assume the mining power of $\alpha$, $b_1$ and $b_2$ is 0.36, 0.29 and 0.27 respectively. The RER of $b_1$ and $b_2$ in four cases is presented in Table 2. For each target $b_i$, their local optimal strategy is to accept the bribes. However, they will suffer losses, comparing with denying the bribes. Furthermore, for all targets, their global optimal strategy is to deny the bribes.

Table 2: RER of target ($RER_{b_i}^{BSSM,H}$ and $RER_{b_i}^{BSSM',H}$). $(x, y)$ indicate the RER of target $b_1$ and target $b_2$ respectively.

| Target $b_2$ \ Target $b_1$ | Accept at bribery initiation stage | Deny at bribery initiation stage |
|---|---|---|
| Accept at bribery initiation stage | (-0.3746%, -0.9311%) | (-6.5856%, 6.4331%) |
| Deny at bribery initiation stage | (8.9069%, -6.6833%) | (3.1083%, 1.1604%) |

## 7.2 The "Bribery Miner's Dilemma" in $BSM$

Similarly, in $BSM$, we consider two targets $b_1$ and $b_2$ with mining power $\beta_1^b$ and $\beta_2^b$. We set $\alpha = 0.3$, $\beta_1^b = 0.2$ and $\varepsilon = 0.02$. We define target $b_i$'s winning condition is to obtain a higher reward than honest mining (i.e., $RER_{b_i}^{BSM,H} > 0$). We calculate the rewards of $b_1$, $b_2$, $a$ and $o$ respectively in four cases. Figure 16 shows the RER and winning conditions for each target in terms of $\beta_2^b$ and $\gamma$.

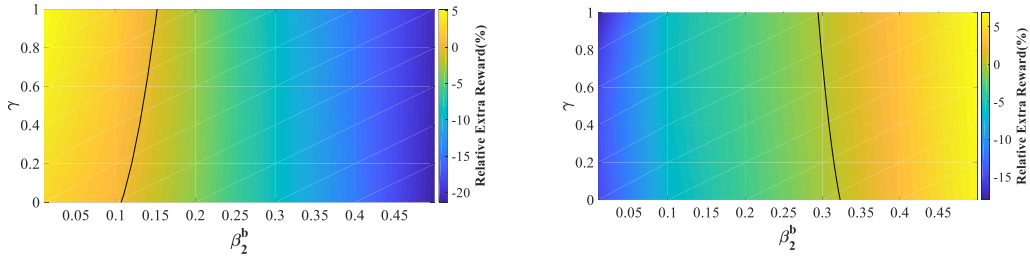

(Fig 16-a) $RER_{b_1}^{BSM,H}$ and winning condition  (Fig 16-b) $RER_{b_2}^{BSM,H}$ and winning condition

Figure 16: RER and winning conditions in $BSM$

The left side of the solid line in Fig 16-a represents the winning condition of target $b_1$, and the right side of solid line in Fig 16-b represents the winning condition of target $b_2$. Fig 16-a indicates that target $b_1$ will obtain extra reward while $b_2$ will suffer losses when $\beta_2^b$ is relatively small. Fig 16-b indicates that target $b_2$ will obtain extra reward while $b_1$ will suffer losses when $\beta_2^b$ is relatively large. The area between two solid lines represents both $b_1$ and $b_2$ will suffer losses (i.e., they will encounter the "bribery miner's dilemma").

We use a more intuitive example to demonstrate the "briery miner's dilemma" in $BSM$. We set $\gamma = 0$ and $\varepsilon = 0.02$, and assume the mining power of $\alpha$, $b_1$ and $b_2$ is 0.36, 0.29 and 0.27 respectively. The RER of $b_1$ and $b_2$ in four cases is presented in Table 3. For each target $b_i$, their local optimal strategy is to accept the bribes. However, they will suffer losses, comparing with denying the bribes. In general, for all targets, their global optimal strategy is to deny the bribes.

Table 3: RER of target ($RER_{b_i}^{BSM,H}$ and $RER_{b_i}^{BSM',H}$). $(x, y)$ indicate the RER of target $b_1$ and target $b_2$ respectively.

| Target $b_2$ \ Target $b_1$ | Accept at bribery initiation stage | Deny at bribery initiation stage |
|---|---|---|
| Accept at bribery initiation stage | (-2.9791%, -2.5061%) | (-9.1528%, 10.2771%) |
| Deny at bribery initiation stage | (7.8802%, -8.5238%) | (1.5172%, 4.0608%) |

## 8 DISCUSSION
### 8.1 Bribery Mining Countermeasure

We present three countermeasures against bribery mining. First, once forking occurs, miners are supposed to choose the branch that they first detect, while ignoring the branch with conflicting transactions. For instance, when a miner detects the transaction $T^A$, and then a fork with two branches occurs (containing transaction $T^A$ and $T^B$ respectively), he should expand the branch with $T^A$. If all miners follow this mining strategy, bribery mining can be avoided effectively. However, this assumption is not realistic as miners can be selfish (they might choose another branch with $T^B$ to obtain higher reward). Nevertheless, the more miners choose to follow this mining strategy, the $\gamma$ smaller, which indicate less rewards for adversaries (the winning probability of adversary's branch decreases).

Secondly, if the victims discover bribery mining in the system, they may be willing to spend money on counter-bribery to win in the forking competition. In general, any miner who obtains reward on the main chain rather than on the adversary's branch can adopt counter-bribery strategies. Meanwhile, the victims would spend no more than the full value of the transaction $T^A$ to implement counter-bribery measures. Once the adversary wins in the competition, the victims will lose the full value of $T^A$. Therefore, the adversaries have to pay the target $b$ higher bribes, which makes bribery mining unprofitable.

Finally, there are potential changes in the role of each miner or pool. The roles of $a$, $o$, $b$ will constantly change over time. Any miner who obtains short-term profits as the adversary at the current moment may also suffer losses as the victim in the future. Therefore, if short-term bribery reward will harm miners' long-term profit potential, they should be motivated not to accept the bribes.

### 8.2 Dynamic Mining Strategy

In the Bitcoin system, miners may adopt various mining strategies to increase their profits. It is difficult to predict the optimal mining strategy, but we can calculate the RER of each mining strategy. For instance, miners with smaller mining power have an advantage in adopting honest mining, comparing with selfish mining. Therefore, rational miners may dynamically adjust their mining strategies in different cases.

## 9 CONCLUSION

We demonstrate that in PoW-based blockchain cryptocurrencies such as Bitcoin, mining attacks can be combined with bribery mining to further expand malicious mining strategies. In $BSSM$, adversaries can obtain relative extra reward of 60% more than honest mining and increase the chain growth rate compared to selfish mining. In $BSM$, adversaries can gain 2% relative extra reward than selfish mining. Both of them will make the targets suffer from the "target miner's dilemma". For each target, their local optimal strategy is to accept the bribes. However, they will suffer losses, comparing with denying the bribes. Furthermore, for all targets, their global optimal strategy is to deny the bribes. Quantitative analysis and simulation have been verified our theoretical analysis. We propose practical measures to mitigate more advanced mining attack strategies based on bribery mining, and provide new ideas for addressing bribery mining attacks in the future. However, how to completely and effectively prevent these attacks is still needed on further research.

## A  State Transitions of BSSM

This section explains Figure 4 in detail.

1. For state 0: **(1)** when $a_s$ finds a valid block, he will reserve it and the length of private chain adds one, which brings the system to state 1 with probability $(1-\rho)\alpha$; **(2)** when $a_i$, $o$ or $b$ finds a valid block, he will publish it on public chain, which brings the system to state 0 with probability $(1-\alpha+\rho\alpha)$.

2. For state $0'_0$: **(1)** when $a_s$ finds a valid block, he will publish it on private chain at once (private branch wins), which brings the system to state 0 with probability $(1-\rho)\alpha$; **(2)** when $o$ finds a valid block, he will publish it on public chain with probability $(1-\gamma)(1-\alpha-\beta^b)$ (public branch wins), or publish it on private chain with probability $\gamma(1-\alpha-\beta^b)$ (private branch wins), which brings the system to state 0; **(3)** when $a_i$ finds a valid block, he will publish it on public chain (public branch wins), which brings the system to state 0 with probability $\rho\alpha$; **(4)** when $b$ finds a valid block, he will choose to accept the bribes and publish it on private chain (private branch wins) with probability $\beta^b$, or choose to deny the bribes and publish it on public chain (public branch wins) with probability $\beta^b$, which brings the system to state 0.

3. For state $0'_b$: **(1)** when $a_s$ finds a valid block, he will publish it on private chain at once (private branch wins), which brings the system to state 0 with probability $(1-\rho)\alpha$; **(2)** when $o$ finds a valid block, he will publish it on public chain (public branch wins) with probability $(1-\gamma)(1-\alpha-\beta^b)$, or publish it on private chain (private branch wins) with probability $\gamma(1-\alpha-\beta^b)$, which brings the system to state 0; **(3)** when $a_i$ finds a valid block, he will publish it on public chain (public branch wins), which brings the system to state 0 with probability $\rho\alpha$; **(4)** when $b$ finds a valid block, he will deny the bribes and publish it on public chain (public branch wins), which brings the system to state 0 with probability $\beta^b$.

4. For state $0'_a$: **(1)** when $a_s$ finds a valid block, he will publish it on private chain at once (private branch wins), which brings the system to state 0 with probability $(1-\rho)\alpha$; **(2)** when $o$ finds a valid block, he will publish it on public chain (public branch wins), which brings the system to state 0 with probability $(1-\alpha-\beta^b)$; **(3)** when $a_i$ finds a valid block, he will publish it on public chain (public branch wins), which brings the system to state 0 with probability $\rho\alpha$; **(4)** when $b$ finds a valid block, he will choose to accept the bribes and publish it on private chain (private branch wins) with probability $\beta^b$, or choose to deny the bribes and publish it on public chain (public branch wins) with probability $\beta^b$, which brings the system to state 0.

5. For state 1: **(1)** when $a_s$ finds a valid block, he will reserve it and the length of private chain adds one, which brings the system to state 2 with probability $(1-\rho)\alpha$; **(2)** when $o$ finds a valid block, he will publish it on public chain, while $a_s$ will publish a reserved block at once and the length of private chain reduces one, which brings the system to state $0'_0$ with probability $(1-\alpha-\beta^b)$; **(3)** when $a_i$ finds a valid block, he will publish it on public chain, which brings the system to state $0'_a$ with probability $\rho\alpha$; **(4)** when $b$ finds a valid blocks, he will publish it on public chain, while $a_s$ will publish a reserved block at once and the length of private chain reduces one, which brings the system to state $0'_b$ with probability $\beta^b$.

6. For state 2: **(1)** when $a_s$ finds a valid block, he will reserve it and the length of private chain adds one, which brings the system to state 3 with probability $(1-\rho)\alpha$; **(2)** when $o$ or $b$ finds a valid block, he will publish it on public chain, while $a_s$ will publish two reserved blocks at once and the length of private chain reduces two (private branch wins), which brings the system to state 0 with probability $(1-\alpha)$; **(3)** when $a_i$ finds a valid block, he will publish it on public chain, which brings the system to state $1'$ with probability $\rho\alpha$.

7. For states $k(k \geq 3)$: **(1)** when $a_s$ finds a valid block, he will reserve it and the length of private chain adds one, which brings the system to state $(k+1)$ with probability $(1-\rho)\alpha$; **(2)** when $o$ or $b$ finds a valid block, he will publish it on public chain, while $a_s$ will publish a reserved block at once and the length of private chain reduces one, which brings the system to state $(k-1)$ with probability $(1-\alpha)$; **(3)** when $a_i$ finds a valid block, he will publish it on public chain, which brings the system to state $(k-1)'$ with probability $\rho\alpha$.

8. For state $1'$: **(1)** when $a_s$ finds a valid block, he will reserve it and the length of private chain reduces one, which brings the system to state $2'$ with probability $(1-\rho)\alpha$; **(2)** when $o$ finds a valid block, he will publish it on public chain, while $a_s$ will publish two reserved blocks at once and the length of private chain reduces two, which brings the system to state $0'_0$ with probability $(1-\alpha-\beta^b)$; **(3)** when $a_i$ find a valid block, he will publish it on public chain, while $a_s$ will publish two reserved blocks at once and the length of private chain reduces two, which brings the system to state $0'_a$ with probability $\rho\alpha$; **(4)** when $b$ finds a valid block, he will publish it on public chain, while $a_s$ will publish two reserved blocks at once and the length of private chain reduces two, which brings the system to state $0'_b$ with probability $\beta^b$.

9. For state $2'$: **(1)** when $a_s$ finds a valid block, he will reserve it and the length of private chain adds one, which brings the system to state $3'$ with probability $(1-\rho)\alpha$; **(2)** when $o$ or $b$ finds a valid block, he will publish it on public chain, while $a_s$ will publish three reserved blocks at once and the length of private chain reduces three, which brings the system to state $0$ with probability $(1-\alpha)$; **(3)** when $a_i$ finds a valid block, he will publish it on public chain, while $a_s$ will publish a reserved block at once and the length of private chain reduces one, which brings the system to state $1'$ with probability $\rho\alpha$.

10. For states $k'(k \geq 3)$: **(1)** when $a_s$ finds a valid block, he will reserve it and the length of private chain adds one, which brings the system to state $(k+1)'$ with probability $(1-\rho)\alpha$; **(2)** when $o$ or $b$ finds a valid block, he will publish it on public chain, while $a_s$ will publish two reserved blocks at once and the length of private chain reduces two, which brings the system to state $(k-1)$ with probability $(1-\alpha)$; **(3)** when $a_i$ finds a valid block, he will publish it on public chain, while $a_s$ will publish a reserved block at once and the length of private chain reduces one, which brings the system to state $(k-1)'$ with probability $\rho\alpha$.

## B  State Transitions of *BSM*

This section explains Figure 10 in detail.

1. For state $1$: **(1)** when $a$ finds a valid block, he will reserve it and the length of private chain adds one, which brings the system to state $2$ with probability $\alpha$; **(2)** when $o$ finds a valid block, he will publish it on public chain, while $a$ will publish a reserved block at once and the length of private chain reduces one, which brings the system to state $0'_0$ with probability $(1-\alpha-\beta^b)$. **(3)** when $b$ finds a valid block, he will publish it on public chain, while $a$ will publish a reserved block at once and the length of private chain reduces one, which brings the system to state $0'_b$ with probability $\beta^b$.

2. For state $1'$: **(1)** when $a$ finds a valid block, he will reserve it and the length of private chain adds one, which brings the system to state $2'$ with probability $\alpha$; **(2)** when $o$ finds a valid block, he will publish it on public chain, while $a$ will publish a reserved block at once and the length of private chain reduces one, which brings the system to state $0'_0$ with probability $(1-\alpha-\beta^b)$; **(3)** when $b$ finds a valid block, he will publish it on public chain, while $a$ will publish a reserved block at once and the length of private chain reduces one, which brings the system to state $0'_b$ with probability $\beta^b$.

3. For state $0'_0$: **(1)** when $a$ finds a valid block, he will publish it on private chain at once (private branch wins), which brings the system to state $0$ with probability $\alpha$; **(2)** when $o$ finds a valid block, he will publish it on public chain with probability $(1-\gamma)(1-\alpha-\beta^b)$ (public branch wins), or publish it on private chain with probability $\gamma(1-\alpha-\beta^b)$ (private branch wins), which brings the system to state $0$; **(3)** when $b$ finds a valid block, he will choose to accept the bribes and publish it on private chain (private branch wins) with probability $\beta^b$, or choose to deny the bribes and publish it on public chain (public branch wins) with probability $\beta^b$, which brings the system to state 0.

4. For state $0'_b$: **(1)** when $a$ finds a valid block, he will publish it on private chain at once (private branch wins), which brings the system to state 0 with probability $\alpha$; **(2)** when $o$ finds a valid block, he will publish it on public chain (public branch wins) with probability $(1-\gamma)(1-\alpha-\beta^b)$, or publish it on private chain (private branch wins) with probability $\gamma(1-\alpha-\beta^b)$, which brings the system to state 0; **(3)** when $b$ finds a valid block, he will deny the bribes and publish it on public chain (public branch wins), which brings the system to state 0 with probability $\beta^b$.